\documentstyle[preprint,eqsecnum,aps]{revtex}
\tighten
\begin{document}
\preprint{gr-qc/9403032 / DESY 94-040 / TUW-93-23}
\draft
\title{Kinetic vs. thermal-field-theory approach to
    \\   cosmological perturbations}
\author{Anton K. Rebhan\thanks{On leave of absence from
	Institut f\"ur Theoretische Physik der
         Technischen Universit\"at Wien}}
\address{DESY, Gruppe Theorie,\\
	Notkestra\ss e 85, D-22603 Hamburg, Germany}
\author{Dominik J. Schwarz
        \thanks{e-mail: dschwarz@ecxph.tuwien.ac.at}}
\address{Institut f\"ur Theoretische Physik,
         Technische Universit\"at Wien,\\
         Wiedner Hauptstra\ss e 8-10/136,
         A-1040 Wien, Austria}

\date{March 1994}

\maketitle

\begin{abstract}
A closed set of equations for the evolution of linear
perturbations of homogeneous, isotropic cosmological models can
be obtained in various ways. The simplest approach is to assume
a macroscopic equation of state, e.g.\ that of a perfect fluid.
For a more refined description of the early universe, a
microscopic treatment is required.  The purpose of this paper is
to compare the approach based on classical kinetic theory to the
more recent thermal-field-theory approach.  It is shown that in
the high-temperature limit the latter describes cosmological
perturbations supported by collisionless, massless matter,
wherein it is equivalent to the kinetic theory approach. The
dependence of the perturbations in a system of a
collisionless gas and a perfect fluid on the
initial data is discussed in some detail. All singular and
regular solutions are found analytically.
\end{abstract}


\section{Introduction}

Early progenitors of large-scale structure of the universe are
usually provided for by small (linear) perturbations of
otherwise homogeneous and isotropic cosmological models
\cite{Peebles80}. After they have come within the Hubble horizon
of the growing universe and their substratum has turned
non-relativistic, they eventually become large, leading to
gravitational collaps --- the corresponding mass scale being set
by the Jeans mass \cite{Jeans}.

The theory of linear perturbations in a
Friedmann-Lema\^{\i}tre-Robertson-Walker (FLRW) model dates back
to Lifshitz' work in 1946 \cite{Lifshitz}. The metric
perturbations and correspondingly the perturbed energy-momentum
tensor are decomposed into scalar, vector and tensor parts
according to their behavior under spatial coordinate
transformations. In linear theory these parts evolve
independently. Any tensor $t^{\mu}{}_{\nu}$ is divided into
a background part $\tilde{t}^{\mu}{}_{\nu}$ and a perturbation
$\delta t^{\mu}{}_{\nu} := t^{\mu}{}_{\nu} -
\tilde{t}^{\mu}{}_{\nu}$.  The background is given by a FLRW
model. However, the mapping of points $x^{\mu}$ on the physical
manifold to points $\tilde{x}^{\mu}$ on the background is not
unique. Unphysical (gauge) modes can be avoided by using
Bardeen's \cite{Bardeen} gauge-invariant variables for the
metric and matter components. The evolution of the metric and
the matter perturbations is governed by the Einstein equation
\begin{equation}
\label{Einstein}
\tilde{G}^{\mu}{}_{\nu} + \delta G^{\mu}{}_{\nu} = - 8\pi G \left(
\tilde{T}^{\mu}{}_{\nu} + \delta T^{\mu}{}_{\nu} \right) \ ,
\end{equation}
where $G^{\mu}{}_{\nu}$ is the Einstein tensor and $T^{\mu}{}_{\nu}$
the energy-momentum tensor. Useful reviews of gauge-invariant
cosmological perturbation theory are, e.g.,  Refs.~\cite{Kodama}
and \cite{Mukhanov}.

In this paper we are mainly interested in comparing the methods
of determining self-consistently the matter perturbations
$\delta T^{\mu}{}_{\nu}$, needed to solve (\ref{Einstein}).  The
Einstein equation describes the evolution of the metric
perturbations for given matter perturbations. The covariant
conservation of the perturbed energy-momentum tensor follows
from the Bianchi identity. This does not fix $\delta
T^{\mu}{}_{\nu}$ completely; additional input is needed. The
simplest possibility is to assume that both the perturbed and
the unperturbed medium satisfy a simple macroscopic equation of
state, e.g., that of a perfect fluid
\cite{Lifshitz,Mukhanov,Sugiyama}.

More complicated forms of matter in the early universe require
of course a microscopic description.  Such a description  was
first developed by Peebles and Yu \cite{Peebles70} on the basis
of kinetic theory \cite{Ehlers}. There the Boltzmann equation in
seven dimensional phase space $(\tau , {\bf x}, {\bf p})$ with
an ansatz for the collision term is used. $\tau$ denotes the
conformal time, ${\bf x}$ the spatial coordinate and ${\bf p}$
the spatial momentum.  The timelike component of the momentum is
not independent, since the particles are on the mass-shell $p^0
= \sqrt{p^2 + m^2}$ ($p=|{\bf p}|$). The Boltzmann equation
together with Eq.~(\ref{Einstein}) has then usually to be solved
numerically. A lot of work has been accomplished within this
framework
\cite{Kodama,Stewart72,Peebles73,Zakharov79,Vishniac82,Bond83,Kasai}.

A different approach has been formulated more recently by
Kraemmer and one of the present authors \cite{Kraemmer}. In a
completely field-theoretical framework, the matter perturbation
$\delta T^{\mu}{}_{\nu}$ is determined by the (thermal) graviton
self-energy \cite{Rebhan91}, evaluated on the given background.
In contrast to kinetic theory, this framework is fully quantum
theoretical from the beginning, and might turn out to be an
alternative to the still to be formulated quantum kinetic
theory. Up to now, this approach was used to incorporate
collisionless, massless matter \cite{Kraemmer,Schwarz} possibly
mixed with a perfect fluid component \cite{Rebhan92a,Rebhan92b},
which can be described perturbatively by the high-temperature
limit of the underlying field theory. In this limit one may
expect equivalence with classical kinetic theory applied to the
quanta of the field theory. A detailed discussion of the
relation between the thermal-field-theory approach and the one
based on kinetic theory is however still missing and will be the
main purpose of the present paper.  The two approaches are
compared for the case of collisionless, massless
matter, with an arbitrary admixture of a perfect fluid.
We will show that both approaches coincide in the
high-temperature limit of thermal field theory.
The exact solutions first found in the field-theoretical
approach \cite{Kraemmer,Rebhan92a} in the form of power series with
infinite radius of convergence will be discussed in some detail
with regard to their dependence on the initial data. Moreover,
by a generalized power-series ansatz further exact solutions are
obtained that are singular as the initial big-bang singularity
is approached.  (The existence of such solutions has been
established first by Zakharov \cite{Zakharov79} and discussed
further by Vishniac \cite{Vishniac82}.)

The paper is organized as follows. In Sect.~II we introduce the
Einstein-de Sitter background and the gauge-invariant metric
potentials and matter perturbations. The Einstein equations are
written down for these variables and their solutions for a
perfect fluid universe are recapitulated. The kinetic theory is explained
in Sect.~III. We follow and add on to
the approach of Kasai and Tomita
\cite{Kasai}, which defines gauge-invariant distribution
functions. A set of integro-differential equations similar to
that already existing in the literature \cite{Stewart72} is
derived for the most general initial conditions. In
Sect.~IV we prove that the same equations emerge from the
thermal-field-theory approach in the high-temperature limit.
The gauge-invariant equations describing a mixture of a collisionless
gas and an isentropic perfect fluid are derived in Sect.~V, and
their exact solutions in terms of (generalized) power series
(details are given in the Appendix)
are discussed in Sect.~VI.
A summary is given in Sect.~VII.

Throughout the paper we use units $\hbar = c = k_B = 1$.  Our
notation is similar to that of \cite{Bardeen} and \cite{Kasai},
but we do not normalize energies and masses to $\sqrt{8\pi G}$.
It differs only in the definition of the metric potentials for
scalar perturbations and in that we expand all functions into
planar waves instead of spherical harmonics, since we consider
only a spatially flat FLRW model. Greek indices take their
values in the set $\{0,1,2,3\}$ and Latin ones in $\{1,2,3\}$.

\section{Gauge-invariant cosmological perturbations}

Before we define the gauge-invariant variables we introduce the
Einstein-de Sitter background. The invariant line element is
given by
\begin{equation}
ds^2 = S^2(\tau) \left( -d\tau^2 + \delta_{ij}dx^i dx^j \right)
\ ,
\end{equation}
with $\tau$ the conformal time measuring the size of the horizon
in comoving coordinates $R_H$.  The background energy-momentum
tensor has perfect-fluid form,
\begin{equation}
\tilde{T}^{\mu}{}_{\nu} = u^{\mu}u_{\nu} (\tilde{E} + \tilde{P}) + \tilde{P}
\delta^{\mu}_{\nu} \ ,
\end{equation}
where $u^{\mu}u_{\mu} = -1$. We shall concentrate on the
radiation dominated epoch, where $\tilde{E} = 3\tilde{P}$. The
scale dependence of the mean energy density is given by
$\tilde{E}(S) = \tilde{E}(S=1) S^{-4}$, and the evolution of the
cosmic scale factor $S(\tau)$ follows from the Friedmann
equation
\begin{equation}
S(\tau) = \sqrt{{8\pi G \tilde{E}(S=1)\over 3}} \tau \ .
\end{equation}

As already explained in the introduction, the perturbed
tensors are divided into scalar, vector and tensor parts:
\begin{equation}
\label{gT}
\begin{array}{rcl}
\delta g_{\mu\nu} &=& \delta g^S_{\mu\nu} + \delta g^V_{\mu\nu}
+ \delta g^T_{\mu\nu} \nonumber \ , \\
\delta T^{\mu}{}_{\nu} &=& \delta T^{S\mu}{}_{\nu} + \delta
T^{V\mu}{}_{\nu} + \delta T^{T\mu}{}_{\nu} \nonumber \ .
\end{array}
\end{equation}
Both tensors have ten independent components, four of which
belong to the scalar, four to the vector and two to the tensor
part (the latter correspond to the two polarizations of a
gravitational wave).

Since the Einstein-de Sitter background is spatially flat and
depends only on the cosmic scale factor $S(\tau)$, any function
$F(\tau,{\bf x})$ can be expanded into plane waves
$e^{\imath{\bf k}{\bf x}}$. A vector and a tensor valued plane
wave is given by ${\bf w}e^{\imath{\bf k}{\bf x}}$ and
${\sf t}e^{\imath{\bf k}{\bf x}}$, respectively, where we
specify the constant (but $\bf k$-dependent)
vector ${\bf w}$ to be transverse (${\bf
w}\cdot{\bf k} = 0$) and the constant symmetric tensor
${\sf t}$ to be transverse-traceless (${\sf t}\cdot{\bf k}
= 0$ and $\text{tr } {\sf t} = 0$). With this any
symmetric tensor can be uniquely decomposed into its scalar,
vector and tensor part. This is done for each individual mode
proportional to $e^{\imath{\bf k}{\bf x}}$ of Eqs.~(\ref{gT}) as
follows [$\{\mu,\nu\} = \{(0,i),(0,j)\}$]:
\begin{mathletters}
\begin{eqnarray}
\delta g^S_{\mu\nu} &=& S^2 \left( \begin{array}{cc}
                                   -2A & \imath \hat{k}_j B^{(0)} \\
                                   \imath \hat{k}_i B^{(0)} &
                                   2\left[ \delta_{ij} H_L +
                                   \left({1\over 3} \delta_{ij} -
                                   \hat{k}_i\hat{k}_j
                                   \right) H_T^{(0)} \right]
                                   \end{array} \right)
                                   e^{\imath{\bf k}{\bf x}}\ , \\
\delta g^V_{\mu\nu} &=& S^2 \left( \begin{array}{cc}
                                   0 & - w_j B^{(1)} \\
                                   - w_i B^{(1)} & - 2\imath
                                   w_{(i}\hat{k}_{j)} H_T^{(1)}
                                   \end{array} \right)
                                   e^{\imath{\bf k}{\bf x}}\ , \\
\delta g^T_{\mu\nu} &=& S^2 \left( \begin{array}{cc}
                                   0 & 0 \\
                                   0 & 2 t_{ij} H_T^{(2)}
                                   \end{array} \right)
                                   e^{\imath{\bf k}{\bf x}}
\end{eqnarray}
\end{mathletters}
and
\begin{mathletters}
\label{mattvar}
\begin{eqnarray}
\delta T^{S\mu}{}_{\nu} &=& {\tilde{E}\over 3} \left( \begin{array}{cc}
                                     - 3\delta & - 4\imath \hat{k}_j
                                     \left( v^{(0)} - B^{(0)}
                                     \right) \\
                                     4\imath \hat{k}^i v^{(0)} &
                                     \delta^i_j \pi_L
                                     + \left( {1\over 3} \delta^i_j
                                     - \hat{k}^i\hat{k}_j \right) \pi_T^{(0)}
                                     \end{array} \right)
                                   e^{\imath{\bf k}{\bf x}}\ , \\
\delta T^{V\mu}{}_{\nu} &=& {\tilde{E}\over 3} \left( \begin{array}{cc}
                                     0 & 4 w_j \left(
                                     v^{(1)} - B^{(1)} \right) \\
                                     - 4 w^i v^{(1)} &
                                     \imath w^{(i}\hat{k}_{j)}
                                     \pi^{(1)}_T
                                     \end{array} \right)
                                   e^{\imath{\bf k}{\bf x}}\ , \\
\delta T^{T\mu}{}_{\nu} &=& {\tilde{E}\over 3} \left( \begin{array}{cc}
                                     0 & 0 \\
                                     0 & t^i{}_j \pi^{(2)}_T
                                     \end{array} \right)
                                   e^{\imath{\bf k}{\bf x}}\ ,
\end{eqnarray}
where ${\bf \hat{k}} = {\bf k}/k$, with $k=|{\bf k}|$.
\end{mathletters}

The metric components $A(\tau,{ k})$, $B^{(a)}(\tau,{ k})$,
$H_L(\tau,{ k})$, $H_T^{(a)}(\tau,{ k})$ and the
matter variables $\delta(\tau,{ k})$, $v^{(a)}(\tau,{ k})$,
$\pi_L(\tau,{ k})$, $\pi_T^{(a)}(\tau,{ k})$ with $a =
\{0,1\}$ transform non-trivially under gauge transformations,
i.e., coordinate transformations of the perturbed
manifold while keeping the background fixed. Only the tensor components
($a=2$) are gauge-invariant by themselves. The gauge transformations for
scalar modes are
\begin{mathletters}
\label{gaugetr}
\begin{equation}
\label{st}
\begin{array}{rcl}
\bar{\tau} &=& \tau + T(\tau, { k})
e^{\imath{\bf k}{\bf x}}\ ,
\nonumber \\
\bar{x}^i &=& x^i - L^{(0)}(\tau, { k}) \imath \hat{k}^i
e^{\imath{\bf k}{\bf x}} \ , \nonumber
\end{array}
\end{equation}
and
\begin{equation}
\label{vt}
\bar{x}^i = x^i + L^{(1)}(\tau, { k}) w^i
e^{\imath{\bf k}{\bf x}}
\end{equation}
\end{mathletters}
for the vector perturbations. From (\ref{st}) and (\ref{vt})
follows that only two scalar ``potentials'' and one vector
``potential'' of the metric
perturbations are gauge-invariant. A set of such
metric potentials is given by
\begin{mathletters}
\begin{equation}
\begin{array}{rcl}
\Phi &:=& 2 \left[ H_L + {1\over 3} H_T^{(0)} + {1\over x}
(B^{(0)} - H_T^{\prime (0)}) \right] \ , \nonumber \\
\Pi &:=& - \left[ A + H_L + {1\over 3} H_T^{(0)} + (B^{(0)}
- H_T^{\prime (0)})^{\prime} + {2\over x} (B^{(0)} -
H_T^{\prime (0)}) \right] \nonumber
\end{array}
\end{equation}
for the scalar part and by
\begin{equation}
\Psi := B^{(1)} - H_T^{\prime (1)}
\end{equation}
\end{mathletters}
for the vector perturbations. The prime denotes a derivative with
respect to
\begin{equation}
\label{xdef}
x := \tau k.
\end{equation}
The variable $x$ is related to the
number of (half-) wavelengths inside the Hubble horizon
\[
{x\over \pi} = {2\over \lambda} R_H ,
\]
and we shall later adopt it as a normalized time variable.

$\Phi$ and $\Pi$ are translated into the variables used by Bardeen
\cite{Bardeen} by
\begin{eqnarray*}
\Phi_H &=& {1\over 2} \Phi \\
\Phi_A &=& - \Pi - {1\over 2} \Phi \ .
\end{eqnarray*}
In the same manner gauge-invariant matter variables can be
defined. For the scalar perturbations we introduce:
\begin{mathletters}
\begin{equation}
\label{smatter}
\begin{array}{rcl}
\eta &:=& \pi_L - \delta
\\
v_s^{(0)} &:=& v^{(0)} - H_T^{\prime (0)} \\
\epsilon_m &:=& \delta + {4\over x}\left( v^{(0)} - B^{(0)}
                                          \right) \\
\epsilon_g &:=& \epsilon_m - {4\over x} v_s^{(0)} \ .
\end{array}
\end{equation}
$\eta$ is called the entropy perturbation, although it
may not coincide with the perturbation in the
true physical entropy. $v_s^{(0)}$ is the
amplitude of the matter velocity, and is related to the shear $\sigma$
by $\sigma=v_s^{(0)}k/S$
\cite{Bardeen}. $\epsilon_m$ is the density contrast on a
spacelike hypersurface which represents the local rest-frame of
matter everywhere, whereelse $\epsilon_g$ is the density
perturbation on a hypersurface whose normal vector has no shear.
The anisotropic pressure $\pi_T^{(0)}$ is gauge-invariant by
itself.

For vector perturbations the invariant matter variables are
\begin{equation}
\label{vmatter}
v_s^{(1)} := v^{(1)} - H_T^{\prime (1)} =: v_c + \Psi \ ,
\end{equation}
where $v_s^{(1)}$ can be understood as in the scalar case and
$v_c$ is the velocity relative to the normal to the
constant-time hypersurface. $v_c k/S$ is the intrinsic
angular velocity
(vorticity) \cite{Bardeen}. The anisotropic pressure $\pi_T^{(1)}$ is
gauge-invariant as above.
For tensor perturbations the only
physical variables are given by the metric potential
$H_T^{(2)} =: H$ and the anisotropic pressure $\pi_T^{(2)}$,
\end{mathletters}
both already gauge-invariant. (Here we follow the
conventions of Ref.~\cite{Bardeen}, which differ from the definitions
of $\Psi$ and $H$ in Ref.~\cite{Rebhan92a} by a factor of 2.)

{}From the Einstein equations (\ref{Einstein}) relations between
the metric potentials  $\Phi, \Pi, \Psi$, and $H$ and the matter
variables $\eta, \epsilon_m, \epsilon_g, \pi_T^{(a)}, v_s^{(a)}$,
and $v_c$ follow. The details are given in Ref.~\cite{Bardeen}.
For the scalar components
\begin{mathletters}
\label{phipi}
\begin{eqnarray}
\label{em}
{x^2\over 3}\Phi &=& \epsilon_m \\
\label{pit}
x^2 \Pi &=& \pi_T^{(0)}\ ,
\end{eqnarray}
\end{mathletters}
which shows the reason for our choice of the metric potentials:
$\Phi$
is a potential for the density perturbation and $\Pi$ the
potential for the anisotropic pressure perturbation. Another set
of equations is useful:
\begin{mathletters}
\label{scalar}
\begin{equation}
\label{trace}
x^2\left( \Phi^{\prime\prime} + {4\over x} \Phi^{\prime} +
{1\over 3}\Phi + {2\over x} \Pi^{\prime} - {2\over 3} \Pi
\right) = - \eta
\end{equation}
\begin{equation}
\label{00}
(x^2 + 3)\Phi + 3x\Phi^{\prime} + 6\Pi = 3\epsilon_g \ .
\end{equation}
\end{mathletters}
Covariant conservation of the perturbed energy-momentum tensor implies
\begin{equation}
\label{scon}
v_s^{\prime (0)} + {1\over x} v_s^{(0)} = {1\over 4} \left(
\epsilon_m + \eta - {2\over 3} \pi_T^{(0)} \right) - {1\over 2}
\Phi - \Pi.
\end{equation}

For vector perturbations the Einstein equations read
\begin{equation}
\label{vector}
{x^2\over 8} \Psi = v_c
\end{equation}
and conservation of the energy-momentum tensor yields
\begin{equation}
\label{vcon}
v_c^{\prime} = - {1\over 8} \pi_T^{(1)}.
\end{equation}

Tensor perturbations fulfill
\begin{equation}
\label{tensor}
x^2 \left( H^{\prime\prime} + {2\over x} H^{\prime} + H \right)
= \pi_T^{(2)} \ .
\end{equation}

Our task in the next Sec.\ (kinetic theory) and in Sec.~IV
(thermal field theory) is to self-consistently
determine the matter perturbations from
a microscopic theory. But the simplest case is to close
the above equations by assuming
macroscopic equations of state.
We end this Section by briefly recapitulating the solutions
obtained with a perfect fluid.

In the case of a perfect fluid the anisotropic pressure
vanishes ($\pi_T^{(a)} \equiv 0$
for all $a$).  Additionally we assume
adiabatic (isentropic) perturbations, i.e., $\eta = 0$.
Eq.~(\ref{phipi}) implies $\Pi \equiv 0$,
and Eq.~(\ref{trace}) reduces to an
ordinary differential equation for the metric
potential $\Phi$. The solution is given
by a spherical Bessel function for regular behavior at
$x=0$ and a spherical von Neumann function for
singular behavior in the origin:
\[
\Phi = {1\over x} \left( c^{(0)} j_1({x\over\sqrt{3}}) +
d^{(0)} y_1({x\over\sqrt{3}}) \right) \ ,
\]
where $c^{(0)}$ and $d^{(0)}$ are constants. The regular solution
approaches a constant for small $x$ (superhorizon scales), which
leads to a growing
$\epsilon_m\sim x^2$ because of Eq.~(\ref{phipi}).
Vector perturbations in a perfect fluid have
constant $v_c$ from (\ref{vcon}),
which is a consequence of the
Helmholz-Kelvin circulation theorem \cite{Zeldovich}.
Regular behavior of $\Psi$ at $x=0$ however
implies $v_c = 0$, for otherwise, $\Psi$ would be proportional to
$1/x^2$. The tensor perturbations are freely
propagating gravity waves; the solution of (\ref{tensor}) is
given by:
\[
H = c^{(2)} j_0(x) + d^{(2)} y_0(x) \ ,
\]
which can be regular or singular as $x\to0$. Their amplitude decays
($\sim x^{-1}$) for large $x$ in accordance with
energy conservation \cite{Zeldovich}.

\section{Kinetic theory approach}

The framework for kinetic theory within general relativity was
set up long ago, see, {\it e.g.},
Refs.~\cite{Ehlers,Linquist}; a recent book
on this subject is Ref.~\cite{Bernstein}. In the context of
cosmological perturbations kinetic theory was first used by Peebels and Yu
\cite{Peebles70}. They discussed the coupled Boltzmann and Einstein
equations for photons interacting with electrons via
Thomson-scattering. For collisionless, massless neutrinos Stewart
\cite{Stewart72} derived the full set of Einstein-Vlasov
equations.
An extensive investigation of density perturbations with
massive and massless neutrinos, radiation and other matter was
performed numerically by Bond and Szalay \cite{Bond83}.

Kinetic theory is applicable for particles whose
de Broglie wavelength is smaller than their mean free path. For
collisionless matter this is certainly the case. Moreover, the
de Broglie wavelength $2\pi/p$ has to be smaller than the Hubble
horizon $R_H = \tau \sim (GT^4)^{-1/2}$, {\it i.e.}, with
\begin{equation}\label{tllmpl}
p^{-1} \sim T^{-1} \ll m_{PL} T^{-2} \ ,
\end{equation}
the temperature $T$ has to be below the Planck scale \cite{Bernstein}.

In this Section we shall concentrate on the case of collisionless, massless
matter; the results for the more general and more realistic case
of a two-component system including a perfect fluid will be
given later on.
The massless case is somewhat
exceptional in that it is rigorously consistent with a thermal
equilibrium situation.  A thermal equilibrium in an expanding
universe can only be obtained in the case of
massless particles, since the
FLRW models provide a conformal timelike Killing-vector only
\cite{Ehlers}, or in the nonrelativistic limit.

In setting up the kinetic approach to cosmological perturbations,
we follow the paper by Kasai and Tomita \cite{Kasai}. They use
gauge-invariant distribution functions \cite{Durrer},
split up into scalar, vector and tensor modes.
Other authors are usually working in a specific
gauge \cite{Stewart72,Peebles73,Zakharov79,Vishniac82,Bond83}.
An infinite set of gauge-invariant equations for the fluid variables
describing density perturbations
has been obtained in Ref.\ \cite{Schaefer}.
The authors of Ref.~\cite{Kasai} use kinetic theory within the tetrad
description of general relativity \cite{Linquist}.
The tetrad is defined by $g_{\mu\nu} = e_{\mu}^a e_{\nu}^b \eta_{ab}$.
It spans an orthonormal basis in the tangent space.

The phase space of an ensemble of particles is described by the
four space-time coordinates $x^{\mu}$ and the momentum components
of the particles $p^{\mu} := dx^{\mu}/ds$, $s$ being an affine
parameter.  Since the particles' energy is on the mass-shell
$p^0 = |{\bf p}| =: p$ (for $m=0$), the dimension of the phase
space is seven. An invariant volume of
momentum 3-space is given by $dp^1 dp^2 dp^3 / p =: d^3 p/p$,
where the indices refer to the tetrad basis.
The distribution function
$F(\tau,{\bf x},{\bf p})$ fulfills the Boltzmann equation
\begin{equation}
\label{Boltzmann}
L(F) := \left( p^{\mu} {\partial \over \partial x^{\mu}} -
\Gamma^{\mu}_{\nu\rho} p^{\nu}p^{\rho} {\partial \over \partial p^{\mu}}
\right) F(\tau,{\bf x},{\bf p}) = C[F] \ .
\end{equation}
The collision term $C$ is a functional of $F$ and has usually to
be put in by hand. It is identically zero for collisionless
matter. $\Gamma^{\mu}_{\nu\rho}$ denotes the usual Christoffel
symbol.

The energy-momentum tensor is defined by
\begin{equation}
T^{\mu}{}_{\nu} := \int {d^3 p \over p} p^{\mu} p_{\nu} F \ .
\end{equation}
In the collisionless case covariant conservation of
$T^{\mu}{}_{\nu}$ follows from the Vlasov equation $L(F)=0$. The
perturbations of the energy-momentum tensor are given by
\begin{equation}
\label{FT}
\delta T^{\mu}{}_{\nu} = \int {d^3 p\over p} \left( \tilde{p}^{\mu}
\tilde{p}_{\nu} \delta F + (\delta p^{\mu} \tilde{p}_{\nu} +
\tilde{p}^{\mu} \delta p_{\nu})\tilde{F} \right) \ .
\end{equation}
The perturbations of the particle momenta $\delta
p^{\mu}$ are calculated with help of the inverse tetrad, which reads
\[
e^{\mu}_a = {1\over S} \left( \delta^{\mu}_a - {1 \over 2}
\delta g^{\mu}{}_a \right)
\]
in linear theory. $\delta p^{\mu} = \delta e^{\mu}_a p^a$
relates $\delta p^{\mu}$ to the metric perturbations.  In
(\ref{FT}) the perturbed distribution function is $\delta F = F
- \tilde{F}$, where $\tilde{F} = \tilde{F}(Sp)$ is the
background distribution function. In thermal equilibrium, the
latter is a Bose-Einstein or a Fermi-Dirac distribution
depending on $S(\tau)p$, the energy of the particles in
comoving coordinates, only. Thus
\[
\tilde{F} = {1\over e^{\beta S p} \pm 1}\ ,
\]
with $\beta S(\tau)$ being the inverse temperature.

The decomposition into scalar, vector and tensor perturbations
in the plane wave expansion is given by
\begin{equation}
\delta F(\tau, {\bf x}, {\bf p}) = \sum_{{\bf k}} \left(
f^{(0)}  + f^{(1)} {{\bf p}\cdot{\bf w}\over p}
+ f^{(2)} {{\bf p}^{\, T} \cdot {\sf t} \cdot{\bf p}\over p^2}
\right) e^{\imath{\bf k}{\bf x}} \ .
\end{equation}

With an isotropic background distribution, the $f^{(a)}$ depend on
$\tau$, the absolute values $k$ and $p$, and the cosine $\mu:=
{\bf k}\cdot{\bf p}/ kp$. For a given mode with wave vector $\bf k$,
we write $f^{(a)}=f^{(a)}(x,p,\mu)$.
Like the matter variables introduced in Eq.~(\ref{mattvar}) the
distribution function is not invariant under the gauge
transformations (\ref{gaugetr}).

\subsection{Scalar perturbations}

For scalar perturbations a gauge-invariant distribution
function is given by
\begin{equation}
I^{(0)}(x,p,\mu) := f^{(0)} + p {\partial \tilde{F}\over
\partial p} \left( {1\over x} (B^{(0)} - H_T^{\prime (0)} ) + \imath
\mu ({1\over 2} B^{(0)} - H^{\prime
(0)}_T)\right) \ ,
\end{equation}
or by
\begin{equation}
J(x,p,\mu) := I^{(0)} - {p\over 2}{\partial \tilde{F}\over
\partial p} \Phi \ .
\end{equation}
According to the expansion theorem for Legendre functions,
\begin{equation}
\label{0exp}
I^{(0)}(x,p,\mu) = \sum_{n=0}^{\infty} a_n^{(0)}(x,p) \imath^n P_n(\mu)
\end{equation}
for every $I^{(0)}(\mu) \in L^2[-1,1]$.
The utility of (\ref{0exp}) becomes apparent by calculating
the gauge invariant matter variables $\epsilon_g, v_s^{(0)}$ and
$\pi_T^{(0)}$ from (\ref{smatter}) and (\ref{FT}) (a more detailed derivation
is given in \cite{Kasai}):
\begin{mathletters} \label{a0}
\begin{equation}\label{a00}
\epsilon_g = {4\pi\over \tilde{E}} \int_0^{\infty}\! dp\, p^3 a_0^{(0)}(x,p)
\end{equation}
\begin{equation}\label{a10}
v_s^{(0)} = - {\pi\over \tilde{E}} \int_0^{\infty}\! dp\, p^3 a_1^{(0)}(x,p)
\end{equation}
\begin{equation}\label{a20}
\pi_T^{(0)} = {12\pi\over 5\tilde{E}} \int_0^{\infty}\! dp\, p^3
a_2^{(0)}(x,p) \ .
\end{equation}
Here the orthogonality of the Legendre polynomials has been used.
\end{mathletters}
$\epsilon_m$ follows from $\epsilon_g$ and $v_s^{(0)}$.
$\eta$ vanishes, since $p^{\mu}p_{\mu} = 0$ in (\ref{FT}) and
$\delta(p^{\mu}p_{\mu}) = 0$, as well. Inserting $J$ into
(\ref{Boltzmann}) for scalar perturbations leads to the equation
\begin{equation}
J^{\prime} - {1\over x} p^{a} {\partial J\over \partial
p^{a}} + \imath \mu J = - \imath \mu p {\partial
\tilde{F}\over \partial p} \left( \Phi + \Pi \right) \ ,
\end{equation}
where $a$ takes the values $1,2,3$ and refers to the tetrad basis.
Its solution is
\begin{equation}
J(x,p,\mu) = e^{-\imath\mu (x - x_0)} J(x_0,p,\mu) - \imath \mu
p {\partial \tilde{F}\over \partial p} \int _{x_0}^x d
x^{\prime} \left(\Phi +\Pi\right)(x^{\prime}) e^{-\imath \mu (x
- x^\prime)} \ .
\end{equation}
With
\[
\epsilon_g = {1\over \tilde{E}} \int {d^3p\over p} p^2 J(x,p,\mu) - 2
\Phi \ ,
\]
and
\[
{1\over \tilde{E}} \int_0^{\infty}\! dp\, p^4 {\partial \tilde{F}\over
\partial p} = - {1\over \pi}
\]
the Einstein equations (\ref{scalar}) read:
\begin{mathletters}
\label{clscalar}
\begin{equation}
\label{cltrace}
\Phi^{\prime\prime} + {2\over x} \Phi^{\prime} + \Phi =
{2\over 3} \left(\Phi + \Pi \right) - {2\over x}\left( \Phi+
\Pi\right)^{\prime}
\end{equation}
\begin{equation}
\label{cl00}
(x^2 + 3)\Phi + 3x\Phi^{\prime} =
6 \left( \Phi + \Pi \right) - 12 \int_{x_0}^x d x^{\prime}
j_0(x-x^{\prime}) \left( \Phi + \Pi \right)^{\prime}(x^{\prime})
+ 12 \sum_{n=0}^{\infty} \beta_n^{(0)} j_n(x-x_0) \ .
\end{equation}
\end{mathletters}
The spherical Bessel functions $j_n(x-x_0)$ are the Fourier
transforms of the Legendre polynomials in Eq.\ (\ref{0exp}).
The last term on the r.h.s.\ determines the initial
conditions. The coefficients $\beta_n^{(0)}$ are defined by
\begin{equation}\label{beta0}
\begin{array}{rcl}
\beta_0^{(0)} &:=& {\pi\over \tilde{E}}
\int_0^{\infty}\! dp\, p^3 a_0^{(0)}(x_0,p)
- \Pi(x_0) - {1\over 2} \Phi (x_0) \ , \\
\beta_n^{(0)} &:=& {\pi\over \tilde{E}} \int_0^{\infty}\! dp\, p^3
a_n^{(0)}(x_0,p) \ ,
\qquad n\geq 1 \ .\\
\end{array}
\end{equation}

\subsection{Vector perturbations}

Here we use the gauge-invariant distribution function
\begin{equation}
I^{(1)}(x,p,\mu) := f^{(1)} + {p\over 2} {\partial \tilde{F}\over
\partial p} B^{(1)}
\end{equation}
and expand it, as above, into Legendre polynomials in $\mu$
\begin{equation}
I^{(1)}(x,p,\mu) = \sum_{n=0}^{\infty} a_n^{(1)}(x,p) \imath^n
P_n(\mu) \ .
\end{equation}
In terms of the coefficients $a_n^{(1)}$, the matter variables are
\begin{mathletters}
\begin{equation}
v_c = {\pi \over \tilde{E}} \int_0^{\infty}\! dp\, p^3
\left(a_0^{(1)}(x,p) + {1\over 5}a_2^{(1)}(x,p) \right)\ ,
\end{equation}
\begin{equation}
\pi_T^{(1)} = - {24\pi \over 5 \tilde{E}} \int_0^{\infty}\! dp\, p^3
\left({1\over 3}a_1^{(1)}(x,p) + {1\over 7}a_3^{(1)}(x,p)
\right) \ .
\end{equation}
\end{mathletters}
For vector perturbations Eq.~(\ref{Boltzmann}) reads:
\begin{equation}
I^{\prime (1)} - {1\over x} p^{a} {\partial I^{(1)}\over \partial
p^{a}} + \imath \mu I^{(1)} =
\imath \mu p {\partial
\tilde{F}\over p} \Psi \ .
\end{equation}
Its solution is:
\begin{equation}
I^{(1)}(x,p,\mu) = e^{-\imath\mu (x - x_0)} I^{(1)}(x_0,p,\mu) + \imath \mu
p {\partial \tilde{F}\over \partial p} \int _{x_0}^x d
x^{\prime} \Psi (x^{\prime}) e^{-\imath \mu (x - x^\prime)} \ .
\end{equation}
With this solution inserted into $v_c$, the Einstein eq.\
(\ref{vector}) becomes
\begin{equation}
\label{clvector}
x^2\Psi = -24 \int_{x_0}^x dx^{\prime} {j_2(x-x^{\prime})\over
x-x^{\prime}} \Psi(x^{\prime}) + 12
\sum_{n=0}^{\infty}
\beta_n^{(1)} \left( j_n(x - x_0) + j^{\prime\prime}_n(x -
x_0) \right) \ .
\end{equation}
The initial conditions for the Volterra type integral equation
(\ref{clvector}) are given by the coefficients
\begin{equation}
\beta_n^{(1)} = {\pi\over \tilde{E}} \int_0^{\infty}\! dp\, p^3
a_n^{(1)}(x_0,p) \ .
\end{equation}

\subsection{Tensor perturbations}

For tensor perturbations the distribution function $f^{(2)}$ is
gauge-invariant itself. With its expansion into Legendre polynomials
\begin{equation}
f^{(2)}(x,p,\mu) = \sum_{n=0}^{\infty} a_n^{(2)}(x,p) \imath^n
P_n(\mu)
\end{equation}
the anisotropic pressure reads:
\begin{equation}
\pi_T^{(2)} = {24\pi \over 15 \tilde{E}} \int_0^{\infty}\! dp\, p^3
\left(a_0^{(2)}(x,p) + {2\over 7}a_2^{(2)}(x,p) + {1\over 21}a_4^{(2)}(x,p)
\right) \ .
\end{equation}
The Vlasov equation for the tensor perturbations reads:
\begin{equation}
f^{\prime (2)} - {1\over x} p^{a} {\partial f^{(2)}\over \partial
p^{a}} + \imath \mu f^{(2)} = p {\partial
\tilde{F}\over \partial p} H^{\prime} \ .
\end{equation}
It is solved by
\begin{equation}
f^{(2)}(x,p,\mu) = e^{-\imath\mu (x - x_0)} f^{(2)}(x_0,p,\mu) +
p {\partial \tilde{F}\over \partial p} \int _{x_0}^x d
x^{\prime} H^{\prime} (x^{\prime}) e^{-\imath \mu (x -
x^\prime)} \ .
\end{equation}
Inserting this into (\ref{tensor}) yields
\begin{eqnarray}
\label{cltensor}
x^2 \left( H^{\prime\prime} + {2\over x} H^{\prime} + H \right)
&=& - 24 \int_{x_0}^x dx^{\prime} {j_2(x-x^{\prime})\over
(x-x^{\prime})^2} H^{\prime}(x^{\prime})\nonumber\\
&& + 3 \sum_{n=0}^{\infty}
\beta_n^{(2)} \left( j_n(x-x_0) + 2j^{\prime\prime}_n(x-x_0) +
j^{(IV)}_n(x-x_0) \right).
\end{eqnarray}
The coefficients encoding the initial conditions are
\begin{equation}
\beta_n^{(2)} = {\pi\over \tilde{E}} \int_0^{\infty}\! dp\, p^3
a_n^{(2)}(x_0,p) \ .
\end{equation}

\section{Thermal-field-theory approach}

Apart from some notable exceptions \cite{Durrer,DurrerAA,Schaefer}
all results given in the literature within kinetic
theory have been derived by choosing a specific gauge. As was
shown in the preceding section, this dependence on the gauge
may be circumvented by
completing the program lined out by Kasai and Tomita \cite{Kasai}.
As we have seen, this requires a skillful redefinition of the
basic distribution function.

A manifestly gauge-invariant
approach is provided by thermal field theory.
For a system containing
gravity as well as matter at a temperature $T$ below the
Planck scale ($T \ll m_{PL}$), the effective action ({\it i.e.}
including all radiative contributions) $\Gamma[g]$ can be split
in a part describing classical gravity, the Einstein-Hilbert
action $S^G$ itself, and an effective action
$\Gamma[g]^M$ induced by the thermal matter with classical
action $S^M$,
\begin{equation}
\label{Gamma}
\left. \Gamma \right|_{T \ll m_{PL}} = S^G + \Gamma^M \ .
\end{equation}
This is
automatically gauge-invariant provided any gauge fields
are subject to background-covariant gauge conditions.

$\Gamma^M$ does not depend on curvature effects in the
leading-order temperature contribution. In the
radiation dominated regime,
the Ricci tensor is proportional to $GT^4$. Thus for temperatures
below the Planck scale ($T \ll m_{PL}$) curvature effects
are lower order in temperature, i.e.\ $\ll T^2$.

The energy-momentum tensor is given by
\begin{equation}
T^{\mu\nu}(x) := {2 \over \sqrt{-g}} {\delta \Gamma^M \over
\delta g_{\mu\nu}} \ ,
\end{equation}
and self-consistent perturbations thereof have to fulfill
\begin{equation}
\label{cons}
\delta T^{\mu}{}_{\nu}(x) = \int_{x^{\prime}} {\delta
T^{\mu}{}_{\nu}(x) \over \delta g_{\rho\sigma}(x^{\prime})} \delta
g_{\rho\sigma}(x^{\prime}) \ .
\end{equation}
This relates the perturbed matter variables to the thermal graviton
self-energy which is defined by
\begin{equation}
\label{defpi}
\sqrt{-g} \Pi^{\mu\nu\rho\sigma}(x,x^{\prime}) :=
{1\over 2} {\delta (\sqrt{-g} T^{\mu\nu}(x)) \over
\delta g_{\rho\sigma}(x^{\prime})} \ .
\end{equation}

A perturbative expansion in Feynman diagrams is appropriate for
weakly interacting matter. Higher loop orders due to internal
graviton propagators are suppressed by powers of $GT^2\ll1$ for
$T\ll m_{PL}$.
A high-temperature expansion of
$\Gamma^M$ in the sense of $k\ll T$ is appropriate for the study of
cosmological perturbations, because the external scale is set
by the Hubble horizon $R_H=\tau\sim (GT^4)^{-1/2}$, which is
$\gg T^{-1}$ for temperatures well below the Planck scale.

First attempts to calculate the thermal gravity self-energy have
been undertaken in Refs.~\cite{Gross}. A complete calculation
of the leading high-temperature contribution
was given first by one of the present authors in Ref.~\cite{Rebhan91}.
For this only one-loop diagrams without internal graviton lines
need to be computed.
Similar issues have been studied in Refs.~\cite{Frenkel}.
Recently, de Almeida et al.\ \cite{Almeida} have computed the
next-to-leading order contributions
for radiation and bosonic matter in the massive,
but still collisionless case.

By calculating the graviton self-energy from thermal field
theory on the cosmological background, one can obtain the
r.h.s.\ of the Einstein equations (\ref{Einstein}) from (\ref{cons})
using the definition (\ref{defpi}). In the
kinetic theory approach the Boltzmann-Einstein system of
equations has to be solved simultaneously. This can be done
analytically for collisionless, massless matter, but in general
only numerical solutions can be obtained. In the thermal
field theory approach, the corresponding problem is to calculate the
graviton self-energy on a curved background space-time. An explicit
reference to perturbations of distribution functions is completely obviated,
which might turn out to be useful when more than the classical limit
is of interest. A complete evaluation of the thermal
graviton self-energy in curved space is certainly a formidable task.
However, restricting our
attention to collisionless, massless matter again, one can take
advantage of the fact that the high-temperature limit of the effective action
possesses, in addition to diffeomorphism invariance, an invariance
under conformal transformations,
which is expressed
by the Ward identity \cite{Kraemmer,Rebhan92a}
\[
\Pi^{\alpha\beta\gamma}{}_{\gamma} = 2
T^{\alpha\beta} \ .
\]
Since the Einstein-de Sitter background is conformally flat,
the graviton self-energy can be simply calculated by
evaluating $\Pi_{\alpha\beta\gamma\delta}$ on flat space-time and
transferring it to the curved space by
multiplication with cosmic scale factors $S$.
This calculation was done by one of the authors in
Refs.~\cite{Rebhan91,Rebhan92a}. We will not repeat it here, but sketch
the most important steps of its derivation.

The graviton self-energy $\Pi_{\alpha\beta\gamma\delta}$ is
related to the Fourier transformed self-energy in flat
space-time $\bar{\Pi}_{\alpha\beta\gamma\delta}$ by
\begin{equation}
\label{ftpi}
\left. \Pi^{\mu\nu\rho\sigma}(x,x^{\prime})
\right|_{g=S^2\eta} = S^{-2}(\tau)
\int {d^4 k \over (2\pi)^4}
e^{\imath k (x - x^{\prime})} \left.
\bar{\Pi}^{\mu\nu\rho\sigma}(k)
\right|_{\eta} S^{-2}(\tau^{\prime}) \ .
\end{equation}
In the high-temperature limit, its tensorial structure turns out
to be the same for any field theory and is given by
\begin{equation}
\left. \bar{\Pi}^{\mu\nu\rho\sigma} \right|_{g=\eta} =
{1\over 2} I^{\mu\nu\rho\sigma}
- {1\over 2} \left( \delta^{\mu}_{\alpha}
\delta^{(\rho}_{\beta} \eta^{\sigma)\nu}
+ \delta^{\nu}_{\alpha} \delta^{(\rho}_{\beta} \eta^{\sigma)\mu}
\right) \eta_{\gamma\delta} I^{\alpha\beta\gamma\delta} \ ,
\end{equation}
which follows in a straightforward calculation from the Feynman
rules of the imaginary time formulation (see  e.g.\
\cite{Landsman}) of thermal field theory. The totally symmetric
quantity $I^{\alpha\beta\gamma\delta}$ is given by
\begin{equation}
I^{\alpha\beta\gamma\delta}(k) = T\sum_{n={p_0\over2\pi iT}} \int
 {d^3{\bf p}\over (2\pi)^3}
{p^{\alpha}p^{\beta}p^{\gamma}p^{\delta}\over p^2 (p-k)^2} \ .
\end{equation}
Sum and integral are conveniently
performed with help of the so-called
Dzyaloshinski algorithm \cite{Dzyaloshinski,Landsman}. Its
values are tabulated in the appendix of Ref.\ \cite{Rebhan91}.
They are polynomials in $k_0/k$ times
\[
Q_0({k_0\over k}) = {1\over 2} \ln\left( {k_0 + k \over k_0 - k}
\right) \ ,
\]
the zeroth Legendre function of second kind. These may be
rewritten with help of the formula
\[
Q_n(z) = P_n(z)Q_0(z) - \sum_{m=1}^n {1\over m} P_{m-1}(z)P_{n-m}(z)
\]
in terms of higher Legendre functions of second kind plus polynomials.

Like all other quantities, the graviton self-energy is expanded
into plane waves. Therefore, only the Fourier transformation of
$k_0$ into $\tau$ in (\ref{ftpi}) has to be performed. The
integrals involved are
\[
\lim_{\gamma \to 0^+} {1\over 2\pi} \int_{-\infty + \imath
\gamma}^{\infty + \imath \gamma} d \omega e^{-\imath \omega
(x-x^{\prime})} Q_n(\omega) = (- \imath)^{n+1}
\Theta(x-x^{\prime}) j_n(x-x^{\prime}) \ ,
\]
where $\omega = k_0/k$ and retarded boundary conditions are
imposed.

In Eq.~(\ref{cons}) this leads to convolution integrals of the form
\[
\sum_n \alpha_n \int^{\infty} dx^{\prime}
\Theta(x-x^{\prime})
j_n(x-x^{\prime}) \{ \Phi, \Pi, \Psi, H\}^{\prime}(x^{\prime}) .
\]
There has to be a finite non-negative integration bound $x_0$
because $\tau \ge 0$.
On the other hand, the functional derivative
in equation (\ref{cons}) is usually defined
without restriction. For that reason we have
to put in the initial conditions for $x_0$ by hand. We do this
by the replacement of the metric potentials by
\[
\{\Phi, \Pi, \Psi, H\}^{\prime}(x) \to \{\Phi, \Pi, \Psi,
H\}^{\prime}(x) \Theta(x - x_0) + \sum_{n=0}^{\infty} \gamma_n^{(a)}
\delta^{(n)}(x-x_0) \ .
\]
The infinite sum provides the initial conditions for
an formally infinite-order differential equation, which is obtained
differentiating the convolution integral above
(see \cite{Rebhan92a}). The infinite sum
yields derivatives of spherical  Bessel functions, which can be
rewritten into spherical Bessel functions.

\subsection{Scalar perturbations}
To arrive at the expressions for the matter perturbations in
thermal field theory, we have to relate them to the graviton self-energy
through Eq.~(\ref{cons}). The matter variables on the
r.h.s.\ of Eqs.~(\ref{scalar}) read:
\[
\eta = 0
\]
and
\begin{equation}
\epsilon_g =  2\Phi + 4\Pi - 4 \int_{x_0}^x d x^{\prime}
j_0(x-x^{\prime}) \left( \Phi + \Pi \right)^{\prime}(x^{\prime})
- 4 \sum_{n=0}^{\infty} \gamma_n^{(0)} j_0^{(n)}(x-x_0) \ .
\end{equation}
Inserting this into (\ref{scalar}), the same equations as in the
kinetic approach, Eq.~(\ref{clscalar}), are obtained, except that
the initial conditions are parametrized differently.
The coefficients $\gamma_n^{(0)}$ can be related to the
$\beta_n^{(0)}$ appearing in Eq.~(\ref{clscalar}) through the
formula
\begin{equation}
j^{\prime}_n=\frac1{2n+1}\left( nj_{n-1}-(n+1)j_{n+1} \right),
\end{equation}
by a linear transformation which is non-singular and of
upper-triangular form (albeit infinite-dimensional).
If $\beta_n^{(0)}=0$ for all $n\ge 3$,
which is a frequently adopted simplifying assumption
in the kinetic approach (the so-called 14-moment approximation,
see, e.g., Ref.~\cite{deGroot}),
then this relationship is given by
\begin{eqnarray}
\gamma^{(0)}_0&=& - \beta^{(0)}_0 - \frac12\beta^{(0)}_2,\nonumber\\
\gamma^{(0)}_1&=& \beta^{(0)}_1,\nonumber\\
\gamma^{(0)}_2&=& - \frac32\beta^{(0)}_2.
\end{eqnarray}

\subsection{Vector perturbations}

In a similar manner we obtain for vector perturbations
\begin{equation}
v_c = -3 \int_{x_0}^x dx^{\prime} {j_2(x-x^{\prime})\over
x-x^{\prime}} \Psi(x^{\prime}) + \sum_{n=0}^{\infty}
\gamma_n^{(1)} \left( j_0(x - x_0) + j_2(x -
x_0) \right)^{(n)}
\end{equation}
which together with
(\ref{vector}) reproduces (\ref{clvector}), if the coefficients
$\gamma^{(1)}_n$ are re-expressed in terms of the $\beta^{(1)}_n$
which parametrize the initial conditions in Eq.~(\ref{clvector}).
Because of $j_0+j_0^{\prime\prime}={2\over3}(j_0+j_2)$, the relationship
between the $\gamma^{(1)}$ and the $\beta^{(1)}$ is, apart from an
over-all sign, the same as in the scalar case. Hence, we have
\begin{eqnarray}
\gamma^{(1)}_0&=&\beta^{(1)}_0+\frac12\beta^{(1)}_2,\nonumber\\
\gamma^{(1)}_1&=&-\beta^{(1)}_1,\nonumber\\
\gamma^{(1)}_2&=&\frac32\beta^{(1)}_2.
\end{eqnarray}

\subsection{Tensor perturbations}
In this case
\begin{equation}
\pi^{(2)}_T = - 24 \biggl[\int_{x_0}^x dx^{\prime} {j_2(x-x^{\prime})\over
(x-x^{\prime})^2} H^{\prime}(x^{\prime}) + \sum_{n=0}^{\infty}
\gamma_n^{(2)} ( \frac1{15}j_0+\frac2{21}j_2+\frac1{35}j_4)^{(n)}
(x-x_0) \biggr] \ ,
\end{equation}
which with (\ref{tensor}) leads again to Eq.~(\ref{cltensor}). Because of
$j_0+2j_0^{\prime\prime}+j_0^{(IV)}=8
( \frac1{15}j_0+\frac2{21}j_2+\frac1{35}j_4)$, the
first three coefficients $\gamma^{(2)}_n$ and $\beta^{(2)}_n$
are related by
\begin{eqnarray}
\gamma^{(2)}_0&=&-\beta^{(2)}_0-\frac12\beta^{(2)}_2,\nonumber\\
\gamma^{(2)}_1&=&\beta^{(2)}_1,\nonumber\\
\gamma^{(2)}_2&=&-\frac32\beta^{(2)}_2,
\end{eqnarray}
if all higher coefficients are zero, exactly as in the scalar case.

Summarizing, the
kinetic theory approach in the gauge invariant formulation of
Ref.~\cite{Kasai} and the thermal-field-theory approach in the high
temperatur limit lead to the same equations for
cosmological perturbations of a
Einstein-de Sitter universe
filled with collisionless, massless particles.
There is a difference, however, in the way the initial conditions
are introduced, which may seem somewhat arbitrary in the field theoretical
approach. In fact, the initial condition terms in
Eqs.~(\ref{clscalar}), (\ref{clvector}), and
(\ref{cltensor}) are completely arbitrary functions due to
the von Neumann expansion theorem. We could therefore just have
added the inhomogeneous terms by hand at the end of the calculation,
but as we have seen
the way in which we have implemented them there is still a close
relation to the inital moments of the initial distribution function
of the kinetic approach.

\section{Collisionless matter and a perfect fluid}

The universe contains, besides collisionless particles, various
different forms of matter as radiation and baryons. An example for
a collisionless, massless gas is provided by neutrinos after
their decoupling below the electroweak scale. Another example
are background gravitons, left over from the epoch of quantum
gravity. In both examples all other
matter may be described by
a perfect fluid during a certain epoch, while collisions keep it
close to thermal equilibrium.

To model such systems we derive the equations for cosmological
perturbations evolving in a mixture of a collisionless,
massless gas and a perfect fluid. We assume that the gas and the
fluid interact only through gravitational forces, thus
\begin{equation}
\label{tmixed}
T^{\mu}{}_{\nu} = {T_{\rm cll}}^{\mu}{}_{\nu} +
{T_{\rm pf}}^{\mu}{}_{\nu} \ .
\end{equation}
To describe the radiation-dominated epoch of the
universe, we assume that the components of the
background perfect-fluid energy-momentum tensor
$\tilde{T}_{\rm pf}{}^{\mu}{}_{\nu}$ obey the same equation of state
as the collisionless gas components, i.e., $\tilde{E}_{\rm pf} = 3
\tilde{P}_{\rm pf}$. For both perturbed energy-momentum tensors, $\delta
{T_{\rm cll}}^{\mu}{}_{\nu}$ and $\delta {T_{\rm pf}}^{\mu}{}_{\nu}$,
gauge-invariant matter variables are defined as in Sec.~II. We
define the ratio of the collisionless background energy-density
to the total background energy-density to be
\begin{equation}
\alpha := {\tilde{E}_{\rm cll} \over {\tilde{E}_{\rm cll} +
\tilde{E}_{\rm pf}}} \ .
\end{equation}
In order to gain a closed set of gauge-invariant equations we
allow only for adiabatic (isentropic) perturbations, i.e., $\eta
= \alpha \eta_{\rm cll} + (1-\alpha) \eta_{\rm pf} = 0$. Entropy
perturbations would arise, e.g., from phase transitions \cite{Mukhanov},
which lie beyond the scope of this work.

The extension of Bardeens \cite{Bardeen} gauge-invariant
formalism to a multi-fluid universe was given in Refs.\
\cite{Kodama,Abbott}. In connection with a collisionless gas
this was done in Ref.\ \cite{Kasai}. Since we deal only with two
components, fulfilling the same background equation of state,
the more general formalism of these references is not necessary
here. The equations we are looking for have been obtained by one
of the present authors in Ref.\ \cite{Rebhan92a}, but without
including the most general initial conditions. The following
gauge-invariant equations are derived from (\ref{Einstein}) with
(\ref{tmixed}).

\subsection{Scalar perturbations}

To arrive at a closed set of equations for scalar perturbations
we define $\Phi_{\rm pf} + \Phi_{\rm cll} := \Phi$ with help of (\ref{em}),
i.e.,
\begin{equation}
\label{ecllm}
{x^2 \over 3} \Phi_{\rm cll} = \alpha \epsilon_{m\ \rm cll} \ .
\end{equation}
Eq.\ (\ref{trace}) does not change since we deal with adiabatic
perturbations only. With (\ref{trace}), (\ref{ecllm}), and
(\ref{pit}), $\Phi, \Phi_{\rm cll}$, and $\Pi$ can be calculated
from $\epsilon_{m\ \rm cll}$ and $\pi_{T\ \rm cll}^{(0)}$. The
anisotropic pressure of the perfect fluid is zero from its
definition. Within the field-theoretic approach from Eqs.\
(\ref{ecllm}) and (\ref{pit})
\begin{mathletters}
\label{mscalar}
\begin{eqnarray}
\label{mphicll}
{x^2 \over 3} \Phi_{\rm cll} &=& - 2 \alpha \biggr[ - \Phi - 2 \Pi +
2 \int_{x_0}^{x} d x^{\prime} \left(j_0(x-x^{\prime})+\frac3x
j_1(x-x^{\prime}) \right) (\Phi+\Pi)^{\prime}(x^{\prime})
\nonumber \\
&& + 2 \sum_{n=0}^{\infty} \gamma_n^{(0)} (j_0+\frac3x
j_1)^{(n)}(x-x_0) \biggr]
\end{eqnarray}
and
\begin{equation}
\label{mpi}
x^2 \Pi = - 12 \alpha \biggr[ \int_{x_0}^x dx^{\prime} j_2(x-x^{\prime})
(\Phi+\Pi)^{\prime}(x^{\prime}) + \sum_{n=0}^{\infty}
\gamma_n^{(0)} j_2^{(n)}(x-x_0) \biggr]
\end{equation}
follow. In terms of $\beta_n^{(0)}$'s (cf.\ Sec.~III) the
inhomogeneous terms respectively read:
\end{mathletters}
\begin{mathletters}
\label{bg0}
\begin{equation}
4\alpha \sum_{n=0}^{\infty} \beta_n^{(0)}
(j_n - \frac3x j_n^{\prime})(x-x_0)
\end{equation}
and
\begin{equation}
6 \alpha \sum_{n=0}^{\infty} \beta_n^{(0)}
(j_n + 3 j_n^{\prime\prime})(x-x_0) \ .
\end{equation}
\end{mathletters}

\subsection{Vector perturbations}

For vector perturbations (\ref{vector}) reads:
\begin{equation}
\label{mvector}
{x^2 \over 8} \Psi = \alpha \left[ -3 \int_{x_0}^x dx^{\prime}
{j_2(x-x^{\prime})\over x-x^{\prime}} \Psi(x^{\prime}) +
\sum_{n=0}^{\infty}\gamma_n^{(1)}\left(j_0+j_2\right)^{(n)}(x-x_0)
\right] + (1-\alpha) v_{c\ \rm pf} \ .
\end{equation}
The covariant conservation of the energy-momentum tensor
(\ref{vcon}) is valid for the gas and the fluid seperately,
i.e., $v_{c\ \rm pf} = const$. In terms of $\beta_n^{(1)}$ the
infinite sum reads:
\[
\frac32\alpha\sum_{n=0}^{\infty} \beta_n^{(1)}
\left( j_n+j^{\prime\prime}_n \right)(x-x_0) \ .
\]

\subsection{Tensor perturbations}

Since the anisotropic pressure for the perfect fluid component
vanishes, (\ref{tensor}) yields
\begin{eqnarray}
\label{mtensor}
x^2 \left( H^{\prime\prime} + {2\over x} H^{\prime} + H \right)
&=& - 24 \alpha \biggr[ \int_{x_0}^x dx^{\prime} {j_2(x-x^{\prime})\over
(x-x^{\prime})^2} H^{\prime}(x^{\prime}) \nonumber \\
&& + \sum_{n=0}^{\infty} \gamma_n^{(2)}
( \frac1{15}j_0+\frac2{21}j_2+\frac1{35}j_4)^{(n)}(x-x_0)
\biggr] \ .
\end{eqnarray}
In terms of $\beta_n^{(2)}$ the inhomogeneity reads:
\begin{equation}
3\alpha \sum_{n=0}^{\infty} \beta_n^{(2)}
\left( j_n+2j^{\prime\prime}_n+j^{(IV)}_n \right)(x-x_0) \ .
\end{equation}

The relation between the coefficients $\beta^{(a)}$ and $\gamma^{(a)}$
is the same as in the purely collisionless case.

\section{Solutions}

The above integro-differential equations cannot, in general, be
solved by, say, a power series ansatz about $x_0$. Instead various
methods have been employed to solve them approximately: direct
numerical integration \cite{Peebles73,Bond83,DurrerAA};
approximation by
a finite sum of spherical Bessel functions \cite{Stewart72}; or by
a finite system of ordinary differential equations \cite{Schaefer}.

However, for $x_0\to0$, which usually is the most interesting point
to define initial conditions anyway, a generalized power series ansatz
can be solved recursively. Exact regular solutions were found in this
way in Refs.~\cite{Kraemmer,Rebhan92a}, where it turned out that
the power series involved have infinite radius of convergence, and
converge faster than trigonometric functions do. In the following,
this will be generalized to include singular solutions as well as
the most general initial data at $x_0=0$. Some of the more unwieldy
details are relegated to the Appendix.

\begin{mathletters}
If $F(x)$ represents one of the metric potentials $\Phi,\Pi,\Psi,H$,
the general solution with $x_0=0$ turns out to be of the form
\begin{equation}\label{psans}
F(x)=C_1 F_{\rm reg}(x)+C_2 x^\sigma F_{\rm sing}(x)
\end{equation}
with
\begin{equation}
F_{\rm reg,sing}(x)=\sum_{n=0}^\infty c^{\rm reg,sing}_n {x^n\over n!}
\end{equation}
and $C_1, C_2$ arbitrary constants. This form follows from demanding that
the coefficients $\beta^{(a)}_n$ remain finite as $x_0\to0$.
\end{mathletters}

When inserted into the integro-differential equations
(\ref{mscalar},\ref{mvector},\ref{mtensor}) this leads to
solvable recursion relations for the coefficients $c_n$ such that
$c_n=c_n(c_0,\ldots,c_{n-1};\alpha,\{\beta\})$, because the
power series representation of the spherical Bessel functions,
\begin{equation}
j_k(x-x')=\sum_{m=0}^\infty{(-1)^m (x-x')^{2m+k}\over(2m)!(2m+1)(2m+3)
\cdots (2m+2k+1)},
\end{equation}
leads to integrals
\begin{equation}
\label{int}
\int_0^x dx'(x-x')^m x^{\prime \nu}=
{m!\over(1+\nu)\cdots(1+m+\nu)}x^{1+m+\nu},
\end{equation}
where $m$ is a natural number and $\nu$ arbitrary.

In solving these recursion relations it turns out that
the coefficients $\beta^{(a)}_n$ (or, equivalently, $\gamma^{(a)}_n$) in
Eqs.~(\ref{mscalar},\ref{mvector},\ref{mtensor}) either have to be all
zero, which also puts to zero $F_{\rm reg}$ except for $H_{\rm reg}$,
or have to satisfy certain constraints dictated by the values $F_{\rm reg}(0)$.
In each case a singular part in Eq.~(\ref{psans}) exists that can
be superimposed on the regular solutions by choosing $C_2\not=0$.
With $x_0=0$, only the regular part depends on the initial data.
The presence of singular perturbations means that the initial singularity
is no longer approximately one
of FLRW type, but is essentially anisotropic. Linear perturbation
theory then applies only for those values of $x$ where $F$ has become
sufficiently small.

It turns out that the singular behavior for $x\to0$ is given by
\begin{equation}\label{exponent}
\sigma^{(a)}_\pm=-{5\over2}+a
 \pm\sqrt{5-32\alpha\over20},
\end{equation}
with $a=0,1,2$ for $(a)=S,V,T$. If the fraction $\alpha$ of energy
density contained in collisionless matter exceeds $\alpha_{\rm crit.}=
\frac5{32}$, $\sigma$ becomes complex, leading to an essential singularity
at $x=0$ with $F$ oscillating like $\cos([{\rm Im}\sigma]\ln x)$.
This asymptotic behavior has been found previously by Zakharov
\cite{Zakharov79} and examined further by Vishniac
\cite{Vishniac82}.\footnote{The result
of Eq.~(\ref{exponent}) agrees with the asymptotic behavior found in
Ref.~\cite{Zakharov79} for scalar, vector, and tensor
perturbations, but not with Ref.~\cite{Vishniac82}, where
the vector case was reported to have a different critical value $\alpha_{\rm
crit.}$.}
As explained by the latter, the essentially singular,
oscillatory behavior for $x\to0$ arises because the momentum flux carried
by the collisionless particles depends preferentially on the
expansion of the universe in their direction of travel, instead
of the net expansion of the volume, which
leads to an excessive feedback for $\alpha>\alpha_{\rm crit.}$.
Thus in this case, the initial singularity is
of a mixmaster type \cite{Misner}.
It has been shown by Lukash et al. \cite{Lukash}
that collisionless matter provides an efficient means to
disperse the initially strong anisotropy of
such models; the singular
parts of our solutions (\ref{psans}) could therefore
correspond to the later stage of
such a scenario where the linear regime has finally been reached.

\subsection{Scalar perturbations}

Regular solutions for scalar perturbations, where $\Phi(x)$
and $\Pi(x)$ stay finite for $x\to0$, are possible only
when the coefficients $\beta_n^{(0)}$ in Eqs.~(\ref{mscalar}) and (\ref{bg0})
satisfy certain constraints.
Evaluation of Eqs.\ (\ref{mscalar}) and their first derivatives
at $x=x_0=0$ yields restrictions on the initial matter
distribution, i.e.,
\begin{equation}
\label{sbeta}
\beta_1^{(0)} = \beta_2^{(0)} = \beta_3^{(0)} = 0 \ ,
\end{equation}
and relations between the initial values of the metric potentials
\begin{mathletters}
\begin{equation}
\Phi(0) + 2\Pi(0) = -4 \beta^{(0)}_0
\end{equation}
and
\begin{equation}
2\Phi^{\prime}(0) = - \Pi^{\prime}(0) \ .
\end{equation}
The restrictions on $\beta^{(0)}_0$, $\beta^{(0)}_1$, and $\beta^{(0)}_2$
can be understand immediately from the Einstein equations written
in the form of Eqs.~(\ref{pit}) and (\ref{00}) together with
covariant conservation of the energy-momentum tensor, Eq.~(\ref{scon}),
by inserting the definitions in Eqs.~(\ref{a0}) and (\ref{beta0});
the vanishing of $\beta^{(0)}_3$ is a consequence of demanding regularity
of $\Pi$ in Eq.~(\ref{mpi}).
\end{mathletters}

The explicit recursion relations defining $\Phi_{\rm
cll}, \Phi_{\rm pf}$, and $\Pi$ follow from the Eqs.\ (\ref{cltrace})
and (\ref{mscalar}). They are listed in the first part of the Appendix.
The recursion relations determine
the regular parts of all potentials uniquely for a
given set of $\beta_n^{(0)}$'s, respecting Eq.\ (\ref{sbeta})
(or equivalently a set of $\gamma_n^{(0)}$'s, satisfying more
involved conditions).
The singular solutions do not depend on the inhomogeneous terms.
Thus they can be added to any regular solution.
Singular solutions for the metric potentials with
poles of first or second order are forbidden for $x_0=0$ since
the integral (\ref{int}) is then divergent for all negative integers
$\nu$.
For finite $x_0$, however such solutions exist. (An example can
be found in Ref.~\cite{Vishniac82}, Eq.~(40b).)

In Fig.~1, regular solutions for $\epsilon_m$, which coincides with
the energy-density contrast $\delta$ on comoving hypersurfaces,
are given in the purely
collisionless case for three different initial data: one where
only $\beta^{(0)}_0$ is nonvanishing, a second where the first four
coefficients have been put to zero, and a more generic one.
All of them grow according to a power-law on superhorizon scales,
where the second (dashed-line) solution exemplifies that an arbitrarily
high power can be achieved by appropriately contrived initial date;
on sub-horizon scales all solutions undergo damped oscillations
whose amplitude falls off like $1/x$ eventually. The exact matching
of the allowed asymptotic behaviors is however seen to depend strongly
on the form of the initial data. The asymptotic regimes themselves
are determined by rather different physical situations. On superhorizon
scales, $x\ll1$, everything can be viewed as being determined by the global
geometry set up in accordance with the inital matter distribution.
For $x\gg1$, microphysics becomes important, and the damped oscillations
there can be understood as the dispersion of density contrast carried by
the collisionless particles through directional dispersion \cite{Bond83}.
The intermediate growth of the amplitude of the dotted
solution in Fig.~1
after its 5th maximum is due to a nonzero $\beta_{20}^{(0)}$.
This demonstrates that higher momenta in the initial distribution
function
may have considerable effects even at subhorizon scales. \footnote{
The sometimes used 14-moment approximation \cite{Schaefer} therefore
potentially misses important details in the evolution of cosmological
perturbations.}

In Fig.~2, a particular regular solution is shown in the two-component
case with equal energy density in the perfect-fluid and in the
collisionless component. In order that the collisionless component
can be matched smoothly to a perfect-fluid behavior for $x\to0$
(in reality there will be a finite time and therefore a finite
value of $x\neq0$, where the collisionless component decouples),
the initial data have been chosen so that
$\Pi(0)=0$, which makes $\Phi_{\rm pf}(0)=\Phi_{\rm cll}(0)$.
After horizon-crossing, the energy-density contrast that remains
in the perfect-fluid component is seen to
be undamped and to have a smaller
phase velocity, namely $c/\sqrt3$, in contrast to $c$ in the
strongly damped collisionless component. The ratio of $\delta_{\rm cll}$
to $\delta_{\rm pf}$ at horizon crossing depends strongly on the
initial data chosen. In Fig.~2 the collisionless component dominates
over the perfect-fluid component up to fairly large values of $x$;
the solutions presented in Refs.~\cite{Rebhan92b} show the opposite
behavior. \footnote{The author of Ref.~\cite{DurrerAA} claims that
$\pi_T^{(0)}<\epsilon_m$ in general, which is seen to hold
true for certain
initial conditions only.}
This is due to having chosen $\beta^{(0)}_5\neq0$ in
the case of Fig.~2.

In Figs.~3--5, singular solutions are displayed. These do not
depend on the initial matter distribution at $x_0=0$, which
underlines their geometrical nature. In Fig.~3, the energy-density
contrast $\delta$ and the anisotropic pressure $\pi_T$ is plotted for
the purely collisionless case $\alpha=1$. There are now oscillations
also on superhorizon scales as discussed above, which go over to
damped oscillations with phase velocity equal to $c$ on subhorizon
scales, showing the same asymptotic behavior there as did the regular
solutions. In Fig.~4, the singular solutions are plotted for $\alpha=5/32$,
where the perfect-fluid
component is just strong enough to eliminate the superhorizon oscillations.
For smaller values of $\alpha$, there a two essentially different
solutions with different degree of singularity. For $\alpha=1/10$, they
are rendered in Fig.~5. In the two-component cases of Figs.~4 and 5,
the perfect-fluid component turns out to have a regular limit for $x\to0$,
growing until horizon crossing, after which they again turn into undamped
acoustic waves with phase velocity $c/\sqrt3$.

\subsection{Vector perturbations}

The requirement of regularity yields the conditions
\begin{mathletters}
\label{vbeta}
\begin{equation}
\label{vbeta2}
{\alpha\over \alpha -1} \left( \beta_0^{(1)} +
{1\over 5} \beta_2^{(1)} \right) = v_{c\ \rm pf}
\end{equation}
and
\begin{equation}
\beta^{(1)}_1 = - \frac37 \beta^{(1)}_3
\end{equation}
for the vector perturbations. These conditions follow from Eq.\
(\ref{mvector}) and its first derivative at $x=x_0=0$.
The singular solutions may be added as in the scalar case.
\end{mathletters}

For
the perfect-fluid case
$\alpha = 0$ there are no regular solutions as mentioned at the
end of Sec.\ II. However, with collisionless matter, vorticity can be
generated on super-horizon scales, which after horizon-crossing
dies out through directional dispersion. This is shown for various
initial data in Fig.~6 for $\alpha=1/2$.
The growth in $v\sim x^2$ at superhorizon scales
is brought about by $\Psi(x)$ approaching a constant for $x\to0$.
A more rapid growth arises when $\Psi$ itself goes like a positive
power of $x$, which is the case for
$\beta^{(1)}_0=0$ and nonvanishing
higher coefficients. One such example is shown by the dashed line
in Fig.~6. In all these cases there is a genuine production of vorticity,
which in the perfect-fluid case is forbidden by the Helmholtz-Kelvin
circulation theorem \cite{BJTJ}.

With different initial
matter distributions it is also possible to arrange for a
nonvanishing initial vorticity in the perfect-fluid component, which
is compensated on superhorizon scales by an equal
amount with opposite sign in the collisionless component. A net
vorticity then survives on subhorizon scales after the dispersion
of the collisionless part. In Fig.~7 such solutions are plotted
for several initial conditions. This interesting possibility
has been studied
extensively in Refs.~\cite{Rebhan92a,Rebhan92b}, to which we refer
the reader for more details.
Here we just mention that the vorticity in the perfect-fluid component
which by the presence of collisionless matter can be reconciled with
an initial FLRW-type singularity, gives rise to magnetic fields
during the transition to the matter-dominated area,
and the limits set by the anisotropy of the cosmic microwave
background are such that astrophysically interesting
amounts of primordial magnetic fields seem possible \cite{Rebhan92b}.

In Fig.~8, we display the singular vector solutions for
$\alpha=1$, 5/32, and 1/10. For $\alpha>5/32$, there are again
superhorizon oscillations, which cease at $\alpha=5/32$, and for
smaller $\alpha$ two different power-law asymptotics occur.
While $v\to0$ for $x\to0$, the metric potential $\Psi$ in fact
diverges. Hence, these solutions do not correspond to a
Friedmann-type initial singularity, but are essentially anisotropic.

On sub-horizon scales, where the mechanism
of directional dispersion becomes operative,
both the regular and the singular solutions
decay like $1/x^2$ for large $x$.

\subsection{Tensor perturbations}

A necessary condition for the regular solutions is obtained from
Eq.\ (\ref{mtensor}) at $x=x_0=0$, i.e.,
\begin{equation}
\label{tbeta}
\beta_0^{(2)} + {2\over 7} \beta_2^{(2)} + {1\over 21} \beta_4^{(2)}
= 0 \ .
\end{equation}
In contrast to vector and scalar perturbations the initial value
$H(0)$ is not fixed  by the initial matter distribution.
This corresponds to the existence of gravitational waves.
Both polarisations evolve with the same $H(x)$, since there is
no preferred direction. Again the singular solutions can be
added to the regular ones.

In Fig.~9 various regular solutions for the metric potential $H$
are given. The ones that start out as constants at superhorizon
scales are similar to the ones found already in the perfect-fluid
case (see above), the only difference being that the amplitude
drops more strongly
at horizon-crossing. On subhorizon scales, these solutions
become gravitational waves and their decay $\sim 1/x$ is
determined only be the expansion, exactly as in the perfect-fluid
case. A novel type of solution is obtained, however, by putting
$\beta^{(2)}_0=0$. As shown by the dashed curve, this gives
a growing solution on superhorizon scales.

Finally, in Fig.~10 we exhibit also the singular solutions for $H$
for the same set of parameters as in Fig.~8. Again, the spectacular
behavior is restricted to superhorizon scales, upon horizon crossing
these solutions again describe ordinary, albeit primordial,
gravitational waves.

\section{Conclusion}

We have presented a detailed analytical
study of cosmological perturbations
of an Einstein-de Sitter universe filled with collisionless massless
particles and a perfect radiation fluid in terms of the
gauge-invariant variables of Bardeen. A closed set of equations
for the latter has been derived within the recently proposed approach
based on thermal field theory on a curved background. There the
response of matter under metric perturbations is determined
by the thermal graviton two-point function, its leading
terms in a high-temperature expansion describing
ultrarelativistic collisionless matter.
We have found complete
equivalence with classical kinetic theory, when the latter is
recast in terms of gauge invariant quantities along the lines
of Ref.~\cite{Kasai}.
A conceptual advantage of the thermal-field-theory approach
appears to be its purely geometrical character ---
no explicit recourse to the generally gauge-variant
distribution functions of kinetic theory is required.
This might turn out to be useful when going beyond the
case of collisionless matter, an issue that we intend
to explore in a future work.

In extension of the work of Ref.~\cite{Kraemmer,Rebhan92a},
we have given the general solutions of the linear perturbation
equations in terms of generalized power series expansions,
the asymptotic behavior of which has been obtained previously
in Ref.~\cite{Zakharov79}. In a strictly FLRW setting, only
the regular solutions are admitted, but the general solutions
might be of interest when to be matched to earlier, inflationary
epochs, where the initial perturbations may have been generated
quantum-mechanically \cite{Gri}.

With special emphasis on the role of the initial data,
we have also presented a selection of explicit results for
regular and singular solutions for scalar, vector, and tensor
cosmological perturbations. Typically, the regular solutions
exhibit growth on superhorizon scales $\sim x^n$, with $n$
a characteristic integer, which can be increased by special
choices of the initial data. The singular solutions have a
small-$x$ behavior dictated only by the ratio $\alpha$ which
gives the ratio of the energy-density in collisionless matter
over the total one. For $\alpha > 5/32$, there is an essential
singularity for $x\to0$, giving rise to superhorizon oscillations,
whereas for $\alpha \leq 5/32$, all perturbations have a (singular)
power-law behavior for small $x$. After horizon-crossing,
all types of perturbations that are carried by collisionless
matter decay. In the scalar and vector case, this is due to
directional dispersion, whereas tensor perturbations only
decrease according to the expansion of the FRLW universe.
We have found that
the transition between superhorizon and subhorizon regimes
depends rather strongly
on the initial data. Whereas previous investigations have always
considered only the simplest cases, our results
indicate that the effective matching of the asymptotic regimes
can vary appreciably by allowing for more complicated initial
matter distributions.

\acknowledgments

A.K.R. acknowledges a stimulating correspondence with Leonid Grishchuk
on rotational perturbations;
D.J.S. thanks Herbert Nachbagauer for discussions on
thermal field theory and Ed Bertschinger for conversations on
cosmological perturbations.
This work was supported by the Austrian Fonds zur F\"orderung der
wissenschaftlichen Forschung (FWF) under project no.\ P9005-PHY.

\appendix
\section*{}

To solve Eqs.~(\ref{cltrace}), (\ref{mscalar}), (\ref{mvector})
and (\ref{mtensor}) the metric potentials $\Phi, \Phi_{\rm cll},
\Pi, \Psi$, and $H$ are expanded into (generalized) power
series. As mentioned in Sec.~VI regular and singular solutions
can be derived in such a way. Since the latter equations
conserve the parity of the metric potentials, the regular
solutions are calculated for the even and the odd part
seperately.

The recursion relations have been implemented in a
{\it Mathematica} \cite{Wolfram} code to calculate the first 60
coefficients and to plot the solutions in Figs.~1--10.

\subsection{Scalar perturbations}

The metric potentials are expanded into power series:
\begin{mathletters}
\begin{eqnarray}
\Phi &=& \sum_{n=0}^{\infty} (-1)^n {x^{2n}\over (2n)!} \left(
\phi_n^{\rm even} + {x\over(2n+1)} \phi_n^{\rm odd} + x^{\sigma^{(0)}}
\phi_n^{\rm sing} \right) \ , \\
\Pi &=& \sum_{n=0}^{\infty} (-1)^n {x^{2n}\over (2n)!} \left(
\pi_n^{\rm even} + {x\over(2n+1)} \pi_n^{\rm odd} + x^{\sigma^{(0)}}
\pi_n^{\rm sing} \right) \ , \\
\Phi_{\rm cll} &=& \sum_{n=0}^{\infty} (-1)^n {x^{2n}\over (2n)!} \left(
q_n^{\rm even} + {x\over(2n+1)} q_n^{\rm odd} + x^{\sigma^{(0)}}
q_n^{\rm sing} \right) \ .
\end{eqnarray}
\end{mathletters}
For simplicity $e_n := 2(\phi_n + \pi_n)$ (for regular even and
odd solutions and for the singular solution) is defined.
{}From Eq.~(\ref{cltrace}), the relations
\begin{mathletters}
\begin{eqnarray}
3(2n+1)\phi^{\rm even}_n &=& 3 \phi^{\rm even}_0 -
e^{\rm even}_0 - 3 e^{\rm even}_n -
\sum_{l=1}^{n-1} (2l+4) e^{\rm even}_l \ , \\
3(2n+2)\phi^{\rm odd}_n &=& 2(3 \phi^{\rm odd}_0 -
e^{\rm odd}_0) - 3 e^{\rm odd}_n -
\sum_{l=1}^{n-1} (2l+5) e^{\rm odd}_l \ , \\
3(\sigma^{(0)} +2n +1) \phi_n^{\rm sing} &=&
{(2n)!\Gamma(\sigma^{(0)}+2)\over \Gamma(\sigma^{(0)}+2n+1)}
(3 \phi^{\rm sing}_0 - e^{\rm sing}_0) - 3 e^{\rm sing}_n -
\nonumber \\
&-& {(2n)!\over \Gamma(\sigma^{(0)}+2n+1)}
\sum_{l=1}^{n-1} {\Gamma(\sigma^{(0)}+2l+1)(\sigma^{(0)}+2l+4)\over
(2l)!} e_l^{\rm sing}
\end{eqnarray}
\end{mathletters}
determine $\Phi$, with $n\ge1$. For the odd regular solutions
$2\phi_0^{\rm odd} = - e_0^{\rm odd}$ and the singular solutions
obey $(\sigma^{(0)}+1)\phi_0^{\rm sing} = - e_0^{\rm sing}$.
Together with ($n\ge0$)
\begin{mathletters}
\begin{eqnarray}
(2n+2)(2n+1)\pi_n^{\rm even} &=& -6\alpha \sum_{l=0}^{n-1}
{2l+2\over(2l+3)(2l+5)} e_{n-l}^{\rm even} + \nonumber \\
& &+ 6\alpha (-1)^n \left[ (2n+2)! \sum_{l=0}^{n+1}{ (-1)^l
\beta^{(0)}_{2n-2l+2}\over (2l)!! (4n-2l+5)!!} + \right.
\nonumber \\
& &+ \left. 3(2n+4)! \sum_{l=0}^{n+2}{(-1)^l
\beta^{(0)}_{2n-2l+4}\over (2l)!! (4n-2l+9)!!} \right] \ , \\
(2n+3)(2n+2)\pi_n^{\rm odd} &=& -6\alpha \sum_{l=0}^{n}
{2l+2\over(2l+3)(2l+5)} e_{n-l}^{\rm odd} + \nonumber \\
& &+ 6\alpha (-1)^n \left[ (2n+3)! \sum_{l=0}^{n+1}{(-1)^l
\beta^{(0)}_{2n-2l+3}\over (2l)!! (4n-2l+7)!!} + \right.
\nonumber \\
& &\left. + 3(2n+5)! \sum_{l=0}^{n+2}{(-1)^l
\beta^{(0)}_{2n-2l+5}\over (2l)!! (4n-2l+11)!!} \right] \ , \\
(2n+2)(2n+1)\pi_n^{\rm sing} &=& -6\alpha {(2n+2)!\over
\Gamma(\sigma^{(0)}+2n+3)} \sum_{l=0}^{n}
{\Gamma(\sigma^{(0)}+2n-2l+1) (2l+2)\over (2n-2l)!
(2l+3)(2l+5)} e_{n-l}^{\rm sing}
\end{eqnarray}
\end{mathletters}
from Eq.~(\ref{mpi}), $\Phi$ and $\Pi$ can be calculated. The
condition (\ref{sbeta}) has to be satisfied to gain regular solutions.
The collisionless part of $\Phi$, $\Phi_{\rm cll}$, is determined
by Eq.~(\ref{mphicll}). From the constant term of the latter equation
\[
\phi_0^{\rm even} + 2\pi_0^{\rm even} = -4 \beta_0^{(0)}
\]
follows for the initial values.
The higher terms give the relations ($n\ge0$):
\begin{mathletters}
\begin{eqnarray}
(2n+2)(2n+1)q_n^{\rm even} &=& 6\alpha \left[ \phi_{n+1}^{\rm even} -
e_{n+1}^{\rm even} + \sum_{l=0}^n \left({1\over 2l+1} +
{3\over (2l+3)(2n+3)}\right)
e_{n-l+1}^{\rm even} \right] + \nonumber \\
& & + 12\alpha(-1)^n \left[ (2n+2)! \sum_{l=0}^{n+1} {(-1)^l
\beta^{(0)}_{2n-2l+2}\over (2l)!!(4n-2l+5)!!} - \right.
\nonumber \\
& & \left. -3(2n+2)!(2n+4) \sum_{l=0}^{n+2} {(-1)^l
\beta^{(0)}_{2n-2l+4}\over (2l)!!(4n-2l+9)!!} \right] \ , \\
(2n+3)(2n+2)q_n^{\rm odd} &=& 6\alpha \left[ \phi_{n+1}^{\rm odd} -
e_{n+1}^{\rm odd} + \sum_{l=0}^{n+1} \left({1\over 2l+1} +
{3\over (2l+3)(2n+4)}\right)e_{n-l+1}^{\rm odd} \right] +
\nonumber \\
& & + 12\alpha(-1)^n \left[ (2n+3)! \sum_{l=0}^{n+1} {(-1)^l
\beta^{(0)}_{2n-2l+3}\over (2l)!!(4n-2l+7)!!} - \right.
\nonumber \\
& & \left. -3(2n+3)!(2n+5) \sum_{l=0}^{n+2} {(-1)^l
\beta^{(0)}_{2n-2l+5}\over (2l)!!(4n-2l+11)!!} \right] \ , \\
(2n+2)(2n+1)q_n^{\rm sing} &=& 6\alpha \left[ \phi_{n+1}^{\rm sing} -
e_{n+1}^{\rm sing} + {(2n+2)!\over\Gamma(\sigma^{(0)}+2n+3)} \times
\right. \\
\lefteqn{\left. \times
\sum_{l=0}^{n+1} \left({1\over 2l+1} +
{3\over (2l+3)(\sigma^{(0)}+2n+3)}\right)
{\Gamma(\sigma^{(0)}+2n-2l+3)\over(2n-2l+2)!}
e_{n-l+1}^{\rm sing} \right] \ .} \nonumber
\end{eqnarray}
\end{mathletters}
The perfect fluid component is provided by $\Phi_{\rm pf} = \Phi
- \Phi_{\rm cll}$.

\subsection{Vector perturbations}

Accordingly, the metric potential $\Psi$ is split up:
\begin{equation}
\Psi = \sum_{n=0}^{\infty} (-1)^n {x^{2n}\over (2n)!} \left(
\psi_n^{\rm even} + {x\over(2n+1)} \psi_n^{\rm odd} + x^{\sigma^{(1)}}
\psi_n^{\rm sing} \right) \ .
\end{equation}
The regular even solutions are defined by
\begin{mathletters}
\begin{eqnarray}
\left[ (2n+2)(2n+1) + {8\alpha\over5}\right] \psi_n^{\rm even} &=&
- 24\alpha \sum_{l=1}^{n} {\psi_{n-l}^{\rm even}
\over(2l+3)(2l+5)} +
\nonumber \\
&+& 12 \alpha (-1)^n \left[ (2n+2)! \sum_{l=0}^{n+1}
{(-1)^l \beta_{2n-2l+2}^{(1)}\over (2l)!! (4n-2l+5)!!} + \right. \nonumber
\\
&+& \left. (2n+4)! \sum_{l=0}^{n+2}
{(-1)^l \beta_{2n-2l+4}^{(1)}\over (2l)!! (4n-2l+9)!!}
\right] \ ,
\end{eqnarray}
for $n\ge 0$. The solution for the regular odd part is provided
for $n\ge 0$ by
\begin{eqnarray}
\left[ (2n+3)(2n+2) + {8\alpha\over5}\right] \psi_n^{\rm odd} &=&
- 24\alpha \sum_{l=1}^{n} {\psi_{n-l}^{\rm odd}\over(2l+3)(2l+5)} +
\nonumber \\
&+& 12 \alpha (-1)^n \left[ (2n+3)! \sum_{l=0}^{n+1}
{(-1)^l \beta_{2n-2l+3}^{(1)}\over (2l)!! (4n-2l+7)!!} + \right. \nonumber
\\
&+& \left. (2n+5)! \sum_{l=0}^{n+2}
{(-1)^l \beta_{2n-2l+5}^{(1)}\over (2l)!! (4n-2l+11)!!}
\right] \ .
\end{eqnarray}
Additionaly the conditions (\ref{vbeta}) have to be satisfied.
Singular solutions are fixed by $\psi_0^{\rm sing}$ and
\begin{eqnarray}
\lefteqn{\left[n(2n+3) + 2n\sigma^{(1)} \right] \psi_n^{\rm sing} =}
 \nonumber \\
& &- 12\alpha {(2n)!\over\Gamma(\sigma^{(1)} +2n+1)}
\sum_{l=1}^{n} {\Gamma(\sigma^{(1)}+2n-2l+1)\over (2n-2l)!}
{\psi_{n-l}^{\rm sing}
\over(2l+3)(2l+5)} \ ,
\end{eqnarray}
for $n\ge 1$.
\end{mathletters}

\subsection{Tensor perturbations}

For the metric potential $H$ the ansatz
\begin{equation}
H = \sum_{n=0}^{\infty} (-1)^n {x^{2n}\over (2n)!} \left(
h_n^{\rm even} + {x\over(2n+1)} h_n^{\rm odd} + x^{\sigma^{(2)}}
h_n^{\rm sing} \right)
\end{equation}
is made. Regular even solutions are defined by
\begin{mathletters}
\begin{eqnarray}
\left[ (2n+1)2n + {8\alpha\over5}\right] h_n^{\rm even} &=&
2n(2n-1) h_{n-1}^{\rm even}
- 24\alpha \sum_{l=1}^{n-1} {h_{n-l}^{\rm even}
\over(2l+1)(2l+3)(2l+5)} +
\nonumber \\
&+& 3 \alpha (-1)^n \left[ (2n)! \sum_{l=0}^{n}
{(-1)^l \beta_{2n-2l}^{(2)}\over (2l)!! (4n-2l+1)!!} + \right. \nonumber
\\
&+& 2(2n+2)! \sum_{l=0}^{n+1}
{(-1)^l \beta_{2n-2l+2}^{(2)}\over (2l)!! (4n-2l+5)!!} +
\nonumber \\
&+& \left. (2n+4)! \sum_{l=0}^{n+2}
{(-1)^l \beta_{2n-2l+4}^{(2)}\over (2l)!! (4n-2l+9)!!}
\right] \ ,
\end{eqnarray}
for $n\ge 1$. The initial value
$h_0^{\rm even}$ specifies the regular solution together with the
inhomogeneous terms. The latter have to respect the condition
(\ref{tbeta}). The solution for the regular odd part is provided
for $n\ge 0$ by
\begin{eqnarray}
\left[ (2n+2)(2n+1) + {8\alpha\over5}\right] h_n^{\rm odd} &=&
(2n+1)2n h_{n-1}^{\rm odd}
- 24\alpha \sum_{l=1}^{n} {h_{n-l}^{\rm odd}
\over(2l+1)(2l+3)(2l+5)} +
\nonumber \\
&+& 3 \alpha (-1)^n \left[ (2n+1)! \sum_{l=0}^{n}
{(-1)^l \beta_{2n-2l+1}^{(2)}\over (2l)!! (4n-2l+3)!!} + \right. \nonumber
\\
&+& 2(2n+3)! \sum_{l=0}^{n+1}
{(-1)^l \beta_{2n-2l+3}^{(2)}\over (2l)!! (4n-2l+7)!!} +
\nonumber \\
&+& \left. (2n+5)! \sum_{l=0}^{n+2}
{(-1)^l \beta_{2n-2l+5}^{(2)}\over (2l)!! (4n-2l+11)!!}
\right] \ .
\end{eqnarray}
Singular solutions are fixed by $h_0^{\rm sing}$ and
\begin{eqnarray}
\lefteqn{\left[n(2n+1) + 2n\sigma^{(2)} \right] h_n^{\rm sing} =
n(2n-1) h_{n-1}^{\rm sing} - }\nonumber \\
& &- 12\alpha {(2n)!\over\Gamma(\sigma^{(2)} +2n+1)}
\sum_{l=1}^{n} {\Gamma(\sigma^{(2)}+2n-2l+1)\over (2n-2l)!}
{h_{n-l}^{\rm sing}
\over(2l+1)(2l+3)(2l+5)} \ ,
\end{eqnarray}
for $n\ge 1$.
\end{mathletters}

\begin{figure}
\caption{Regular density perturbations
on comoving hypersurfaces in the purely
collisionless case ($\alpha=0$).
The full line shows a regular solution for
$|\delta|=|\epsilon_m|$
with $\beta^{(0)}_0 = -5/28$ and all other $\beta^{(0)}_n$'s
vanishing.  The dashed line represents a solution with
$\beta^{(0)}_5 = -1309/20$ nonvanishing only. For the dotted line
$\beta^{(0)}_0$ takes the above value and $\beta^{(0)}_5 = -10$,
$\beta^{(0)}_{20} = 60$.}
\label{fig1}
\end{figure}

\begin{figure}
\caption{Regular scalar perturbations in a
two-component universe ($\alpha = 1/2$).
The various lines show regular solutions for
$|\delta_{\rm cll}|$ (full line), $|\delta_{\rm pf}|$
(dotted line), and $|\pi^{(0)}_T|$ (dashed line).
The initial conditions
are specified by $\beta_0^{(0)} = -1/4, \beta_1^{(0)} = -1309/20$, and
$\beta_4^{(0)} = 7/8$.}
\label{fig2}
\end{figure}

\begin{figure}
\caption{Singular solutions for $|\delta|$ (full line)
and $|\pi^{(0)}_T|$ (dashed line) for $\alpha=1$
(purely collisionless case). Superhorizon
oscillations show up for $x/\pi<1$. Notice the logarithmic
scale in $x$ --- the solutions are essentially singular as
$x\to0$.
The second solution corresponding to $\sigma_-^{(0)}$ in
Eq.~(\protect{\ref{exponent}}), which is not plotted,
differs essentially by a phase.}
\label{fig3}
\end{figure}

\begin{figure}
\caption{Singular scalar perturbations for the critical value of
$\alpha = 5/32$. $|\delta_{\rm cll}|$
(full line), $|\delta_{\rm pf}|$ (dotted line), and $|\pi_T^{(0)}|$
(dashed line) are plotted.}
\label{fig4}
\end{figure}

\begin{figure}
\caption{Singular scalar perturbations in the
subcritical case ($\alpha < 5/32$). The two
different modes ($\sigma_{\pm}^{(0)}$) are shown by the
full lines for $|\delta_{\rm cll}|$ and by the dotted ones for
$|\delta_{\rm pf}|$. They have different power-law behavior
on superhorizon scales.
}
\label{fig5}
\end{figure}

\begin{figure}
\caption{Regular vector perturbations with
$v_{c\ \rm pf}=0$. The solutions for $|v_c|$ are given with only
$\beta^{(1)}_0=-21/16$ nonzero (full line), $\beta^{(1)}_1 =
-1197/80$ alone (dashed line), and $\beta^{(1)}_0 = -21/16,
\beta^{(1)}_1 = -10$ and $\beta^{(1)}_{10} = 20$ for the
dotted line. }
\label{fig6}
\end{figure}

\begin{figure}
\caption{Regular vector perturbations with non-vanishing
vorticity in the perfect-fluid component.
The contribution of $v_{c\ \rm pf} = -21/16$ to
$|v_c|$ is plotted dashed-dotted. The other lines show various
solutions for $|v_c|$ using the same initial conditions as in
Fig.~\protect{\ref{fig6}} exept for $\beta_2^{(1)}$, which is
determined by Eq.~(\protect{\ref{vbeta2}}).}
\label{fig7}
\end{figure}

\begin{figure}
\caption{Singular vector perturbations in terms of $|v_c|$.
The different lines correspond to different values of $\alpha$
and to different modes. The solutions with, e.g., $\alpha = 1$
(thin full line) show superhorizon oszillations, whereas solutions
for $\alpha_{\rm crit.}$ (thick full line) and
below, e.g., $\alpha = 1/10$ (dashed-dotted and dotted)
do not. The latter two lines correspond to the two different
modes from Eq.\ (\protect{\ref{exponent}}). In the supercritical case
both modes differ by their phase only.}
\label{fig8}
\end{figure}

\begin{figure}
\caption{Regular tensor perturbations.
The solutions, representing gravitational waves for $x\gg1$,
are normalized
to $H(0)=1$, exept for the dashed line ($H(0)=0, \beta^{(2)}_1=63/2$).
The full line shows the solution with all $\beta^{(2)}_n$'s vanishing.
Again the dotted line represents a more general case with
$\beta^{(2)}_1 = 1, \beta^{(2)}_4 = 5, \beta^{(2)}_6 = -10$, and
$\beta^{(2)}_7 = 20$ nonzero.}
\label{fig9}
\end{figure}

\begin{figure}
\caption{Singular tensor perturbations. $|H|$ is plotted for
the same values of $\alpha$ as for the
vector perturbations of Fig.~\protect{\ref{fig8}}.}
\label{fig10}
\end{figure}


\begin{references}

\bibitem{Peebles80}P. J. E. Peebles, {\it The Large-Scale
         Structure of the Universe} (Princeton University Press,
         Princeton, New Jersey, 1980).
\bibitem{Jeans}J. Jeans, Philos. Trans. R. Soc. London A {\bf 199}, 491 (1902).
\bibitem{Lifshitz}E. Lifshitz, Zh. Eksp. Teor. Fiz. {\bf 16}, 587 (1946);
         E. Lifshitz and I. Khalatnikov, Adv. Phys. {\bf 12}, 185 (1963).
\bibitem{Bardeen}J. M. Bardeen, Phys. Rev. D {\bf 22}, 1882 (1980).
\bibitem{Kodama}H. Kodama and M. Sasaki, Prog. Theor. Phys. Suppl.
         {\bf 78}, 1 (1984).
\bibitem{Mukhanov}V. F. Mukhanov, H. A. Feldman, and R. H. Brandenberger,
         Phys. Rep. {\bf 215}, 203 (1992); R. Durrer, {\it Gauge Invariant
         Cosmological Perturbation Theory}, Report No. ZU-TH14/92, 1993.
\bibitem{Sugiyama}N. Sugiyama and N. Gouda, Prog. Theor. Phys. {\bf 88},
         803 (1992);and references therein.
\bibitem{Peebles70}P. J. E. Peebles and J. T. Yu, Astrophys. J. {\bf 162}, 815
(1970).
\bibitem{Ehlers}J. Ehlers, in {\it General Relativity and Gravitation}, edited
         by R. K. Sachs (Academic Press, New York, 1971); J. M. Stewart,
         {\it Non-equilibrium Relativistic Kinetic Theory}, (Springer-Verlag,
         New York, 1971).
\bibitem{Stewart72}J. M. Stewart, Astrophys. J. {\bf 176}, 323 (1972).
\bibitem{Peebles73}P. J. E. Peebles, Astrophys. J. {\bf 180}, 1 (1973).
\bibitem{Zakharov79}A. V. Zakharov, Zh. Eksp. Teor. Fiz. {\bf 77}, 434 (1979).
\bibitem{Vishniac82}E. T. Vishniac, Astrophys. J. {\bf 257}, 456 (1982).
\bibitem{Bond83}J. R. Bond and A. S. Szalay, Astrophys. J. {\bf 274}, 443
(1983).
\bibitem{Kasai}M. Kasai and K. Tomita, Phys. Rev. D {\bf 33}, 1576 (1986).
\bibitem{Kraemmer}U. Kraemmer and A. Rebhan, Phys. Rev. Lett. {\bf 67},
         793 (1991).
\bibitem{Rebhan91}A. Rebhan, Nucl. Phys. {\bf B351}, 706 (1991).
\bibitem{Schwarz}D. J. Schwarz, {\it Analytic Solutions for Cosmological
         Perturbations in Multi-Dimensional Space-Time},
         Technische Universit\"at Wien Report No. TUW-93-07,
         1993 (to be published).
\bibitem{Rebhan92a}A. Rebhan, Nucl. Phys. {\bf B369}, 479 (1992);
	in {\it Relativistic Astrophysics and Cosmology}, S. Gottl\"ober,
	J. P. M\"ucket, and V. M\"uller (eds.), (World Sci., Singapore, 1992)
        p. 137.
\bibitem{Rebhan92b}A. Rebhan, Astrophys. J. {\bf392}, 385 (1992).
\bibitem{Zeldovich}Y. B. Zeldovich and I. D. Novikov, {\it
         Relativistic Astrophysics Vol. 2: The Structure and
         Evolution of the Universe}, (The Univesity of Chicago Press,
         Chicago, 1983).
\bibitem{Linquist}R. W. Linquist, Ann. Phys. (N.Y.) {\bf 37}, 487 (1966).
\bibitem{Bernstein}J. Bernstein, {\it Kinetic Theory in the
         Expanding Universe}, (Cambridge University Press, Cambridge, 1988).
\bibitem{Durrer}A gauge-invariant kinetic approach was developed
         independently from Ref.~\cite{Kasai} in
         R. Durrer and N. Straumann, Helv. Phys. Acta {\bf 61},
         1027 (1988).
\bibitem{Schaefer}R. K. Schaefer, Int. J. Mod. Phys. A {\bf 6},
         2075 (1991).
\bibitem{DurrerAA}R. Durrer, Astron. Astrophys. {\bf 208}, 1 (1989).
\bibitem{Gross}D. J. Gross, M. J. Perry, and L. G. Yaffe, Phys.
         Rev. D {\bf 25}, 330 (1982); Y. Kikuchi, T. Moriya, and
         H. Tsukahara, Phys. Rev. D {\bf 29}, 2220 (1984); P. S.
         Gribosky, J. F. Donoghue, and B. R. Holstein, Ann.
         Phys. (N.Y.) {\bf 190}, 149 (1989).
\bibitem{Frenkel}J. Frenkel and J. C. Taylor, Z. Phys. C {\bf
         49}, 515 (1991); F. T. Brandt, J. Frenkel, and J. C.
         Taylor, Nucl. Phys. {\bf B374}, 169, (1992); F. T.
         Brandt and J. Frenkel, Phys. Rev. D {\bf 47}, 4688 (1992);
         {\it The graviton self-energy in thermal quantum gravity},
         Report No. IFUSP-P-1050, 1993.
\bibitem{Almeida}A. P. de Almeida, F. T. Brandt, and J. Frenkel,
         {\it Thermal matter and radiation in a gravitational
         field}, Report No. IFUSP-P-1070, 1993.
\bibitem{Landsman}N. P. Landsman and Ch. G. van Weert, Phys.
         Rep. {\bf 145}, 141 (1987).
\bibitem{Dzyaloshinski}I. E. Dzyaloshinski, Zh. Eksp. Teor. Fiz. {\bf
         42}, 1126 (1962); R. D. Pisarski, Nucl. Phys. {\bf
         B309}, 476 (1988); J. Frenkel and J. C. Taylor, Nucl. Phys.
         {\bf B334}, 199 (1990).
\bibitem{deGroot}S. R. de Groot, W. A. van Leeuwen, and Ch. G. van Weert,
         {\it Relativistic Kinetic Theory}, (North-Holland Publishing
         Company, Amsterdam, 1980).
\bibitem{Abbott}L. F. Abbott and M. B. Wise, Nucl. Phys. {\bf
         B237}, 226 (1984).
\bibitem{Misner}C. W. Misner, Phys. Rev. Lett. {\bf 22}, 1071 (1969);
	V. A. Belinskii, I. M. Khalatnikov and E. M. Lifshitz,
	Adv. Phys. {\bf 19}, 525 (1970).
\bibitem{Lukash}V. N. Lukash, I. D. Novikov and A. A. Starobinskii,
        Sov. Phys. - JETP {\bf 42}, 757 (1976).
\bibitem{BJTJ}B. J. T. Jones, Rev. Mod. Phys. {\bf 48}, 107 (1976).
\bibitem{Gri}L. P. Grishchuk, Sov. Phys. - JETP {\bf 40}, 409 (1975);
	Phys. Rev. D {\bf 48}, 5581 (1993).
\bibitem{Wolfram}S. Wolfram, {\it Mathematica} 2nd Ed., (Addison-Wesley
Publishing Company,
        Redwood City, 1991).
\end{references}
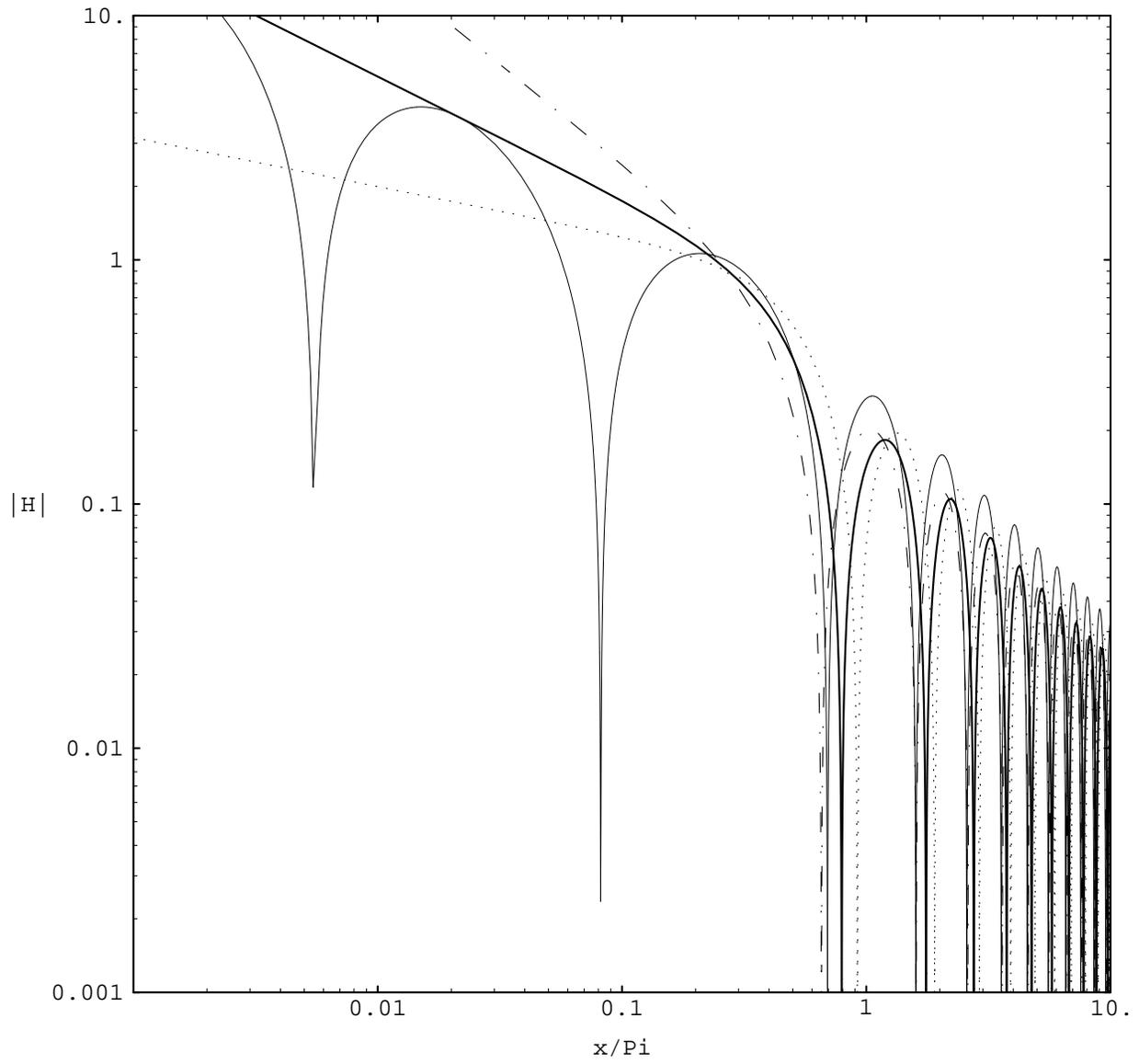
\end{document}